# Observation of collective flatbands in hybrid Mie–grating THz metasurfaces

Jie Ji[1] | Thanh Xuan Hoang[2 †] | Ayan Nussupbekov[3] | Siyan Zhou[4] | Kehan Wang[1] | Yuri Kivshar[5] | Jaime Gómez Rivas[1]

[1]Institute for Complex Molecular Systems and Casimir Institute, Department of Applied Physics and Science Education, Eindhoven University of Technology, Eindhoven, the Netherlands | [2]Institute of High Performance Computing, Agency for Science, Technology and Research (A*STAR), Singapore, Singapore | [3]Institute of Advanced Intelligence and Computing, Agency for Science, Technology and Research (A*STAR), Singapore, Singapore | [4]Department of Electrical and Photonics Engineering, Technical University of Denmark, Kongens Lyngby, Denmark [5]Research School of Physics, Australian National University, Canberra, Australia |

**Correspondence:** Jie Ji (j.ji@tue.nl;) | Yuri Kivshar (yuri.kivshar@anu.edu.au) | Jaime Gómez Rivas (j.gomez.rivas@tue.nl)

## ABSTRACT

We propose a novel type of functional metasurfaces created by a hybridization of Mie resonators with high-index optical gratings. We demonstrate that such hybrid metasurfaces possess flatbands characterized by zero group velocity and strong energy localization. We experimentally observe the flatbands in free-standing terahertz (THz) metasurfaces, where dielectric interconnections between Mie resonators, forming the grating, mediate multipolar hybridization and provide a geometric control over collective resonances. Our THz near-field measurements with a dipole excitation uncover long-lived coherent oscillations associated with the quasi-flat magnetic-dipolar resonant modes, while a spatial mapping resolves the resonant electromagnetic response across the dielectric interconnections. We vary the grating width continuously to tune the hybrid magnetic-dipolar resonances and substantially enhance the quality factor ($Q$ factor), while preserving the quasi-flatband dispersion. Complementary Bloch-resolved Purcell calculations reveal the substantial Purcell enhancement with increasing $Q$ factor, followed by a trend towards saturation and highlighting the interplay between radiative confinement and local emitter-mode coupling. We believe our findings establish the engineering of dielectric interconnections in periodic lattices as a novel strategy for spectrally tuning high-$Q$ flatbands in these metastructures, tailoring localized electromagnetic interactions in functional optical metasurfaces.

## 1 Introduction

Photonic band dispersion plays a fundamental role in governing light propagation, confinement, and electromagnetic interactions in periodic optical structures [1, 2, 3]. Among the various photonic dispersion characteristics, flatbands have attracted considerable attention due to their suppressed dispersion and reduced group velocities, which provide favorable conditions for enhanced light–matter interactions and enable potential applications in various fields, such as spontaneous emission enhancement [4], lasing [5], and polariton condensation [6]. Various strategies have been explored to realize flat photonic bands by engineering lattice geometries and manipulating mode coupling and hybridization [7, 8, 9]. In geometrically frustrated lattices, such as Lieb and kagome lattices, destructive interference among different propagation pathways can suppress wave transport, giving rise to dispersionless bands. Alternatively, flatbands can emerge from engineered coupling and hybridization between optical modes, where their interactions reshape the band structure and suppress dispersion [10, 11]. More recently, moiré photonic structures formed by twisting coupled lattices have emerged as another promising platform for band flattening, in which twist-angle-dependent inter-layer coupling and hybridization can lead to strongly reduced band dispersion [12, 13, 14]. Although these mechanisms can give rise to flat photonic bands with suppressed group velocities, band flatness alone does not necessarily guarantee enhanced light–matter interactions, which also depend critically on the radiative properties and spatial electromagnetic field distributions of the underlying optical modes [15, 16, 17]. In this context, all-dielectric metasurfaces are particularly attractive because their low intrinsic material losses and diverse electromagnetic resonances provide opportunities to engineer high-$Q$ photonic modes [18, 19, 20].

† Present address: Oakville, Ontario, Canada.

Beyond the advantages of low material losses, nonlocal electromagnetic interactions offer an additional degree of freedom for engineering photonic band dispersion. By exploiting spatially extended coupling between resonators, these interactions enable control over the momentum-dependent dispersion and radiative properties of collective modes [21, 22]. Such collective resonances have been widely investigated in periodic photonic structures, including plasmonic metasurfaces [23, 24, 25, 26], photonic crystals [27, 28], and dielectric resonators [29, 30, 31]. Their tunable dispersive and radiative characteristics enable a wide range of functionalities, including analog optical computing [32, 33, 34] and directional light manipulation [35, 36, 37, 38]. Of particular interest are dielectric resonators, which support electric and magnetic Mie resonances with distinct symmetries and coupling characteristics across material platforms, including silicon [39, 40], lithium niobate [41, 42], and the diverse family of two-dimensional materials [43, 44, 45]. The ability to manipulate these multipolar resonances has established dielectric metasurfaces as an effective platform for controlling light at the subwavelength scale. When arranged in periodic lattices, coupling between these resonances gives rise to collective optical modes whose dispersion is governed by the combined effects of individual resonator characteristics and lattice-mediated electromagnetic interactions [46, 47]. Importantly, interference and hybridization among different multipolar channels can reshape collective band structures and, under suitable conditions, suppress their dispersion to form nonlocal flat photonic bands [48, 49]. These interactions also provide additional degrees of freedom for tailoring the spatial electromagnetic field distributions and radiative characteristics of such high-$Q$ modes.

In the THz regime, high-$Q$ quasi-flat photonic bands are particularly attractive for achieving strong resonant field enhancement and narrowband spectral selectivity, offering opportunities for various applications [50, 51], including sensing [52, 53] and next-generation communication [54]. However, simultaneously achieving high-$Q$ resonances, weak band dispersion, and geometric tunability remains challenging, as it requires precise control over the coupling and radiative properties of collective modes. In this work, we demonstrate a free-standing THz Mie–grating metasurface that exploits grating-mediated multipolar hybridization to engineer photonic band structures. By tailoring the coupling between electric and magnetic multipolar resonances, we realize strongly anisotropic dispersions featuring a high-$Q$ quasi-flatband and a Dirac-like cone along orthogonal momentum directions. Phase-resolved THz near-field microscopy reveals long-lived coherent oscillations and spatially resolved local density of optical states (LDOS) enhancement associated with the collective quasi-flat-band mode. Furthermore, we demonstrate that the grating geometry provides control over the modal $Q$-factor and local electromagnetic response, supported by Bloch-resolved Purcell factor calculations. Our findings establish grating-mediated multipolar hybridization as a versatile strategy for engineering collective photonic modes and enhancing local light–matter interactions.

## 2 Results and Discussion

The free-standing Mie–grating metasurface integrates a two-dimensional square lattice of cylindrical dielectric resonators with a one-dimensional dielectric grating that serves as an interconnecting network, forming a hybrid photonic architecture with distinct yet coupled resonant and diffractive responses as shown in Figure 1. The free-standing architecture eliminates substrate-supported optical effects, thereby facilitating the engineering and investigation of multipolar hybridizations. The cylindrical resonator array shown in Figure 1(a) supports electric and magnetic Mie resonances whose lattice-mediated electromagnetic coupling gives rise to collective photonic modes or surface lattice resonances [55, 56]. Beyond providing mechanical support by interconnecting adjacent resonators in the free-standing structure, the dielectric grating in Figure 1(b) acts as an active photonic element that supports waveguide-like Bloch modes and mediates additional electromagnetic coupling between Mie resonators in the resulting hybrid architecture, which is displayed in Figure 1(c). Increasing the grating width progressively transforms the structure from a discrete resonator array toward a continuous grating-like architecture, enabling systematic control over the resonant characteristics and photonic band structure.

To experimentally realize the proposed concept, silicon metasurfaces were fabricated using a high-resistivity silicon wafer with a thickness of 200 $\mu$m. The metasurfaces consist of a square lattice of cylindrical resonators interconnected by a dielectric grating oriented along the $y$-direction and extending to the device boundaries. The cylindrical resonators have a radius of $R = 120$ $\mu$m, and the lattice period is $P = 258$ $\mu$m. To investigate the influence of the grating width, four samples were fabricated with varying widths of $\mathrm{w} = 25$, 50, 75, and 100 $\mu$m, respectively. The scanning electron microscopy image in Figure 1(c,ii) corresponds to the fabricated metasurface with a grating width of $\mathrm{w} = 50$ $\mu$m, confirming the periodic arrangement of the resonators and their interconnected geometry. Additional details of the fabrication process are provided in the Methods section, Supplementary Section S1, and Supplementary Figure S1.

### 2.1 Flatbands induced by multipolar hybridization

The dielectric interconnections are expected to modify the coupling between the collective electric- and magnetic-dipolar resonances in the hybrid Mie–grating metasurface. To elucidate their influence on the resonant response, we first simulated and compared the normal-incidence transmission spectra of the cylindrical resonator array and the hybrid Mie–grating metasurface. Figure 2(a) shows that the cylinder array supports two hybrid eigenmodes arising from lattice-mediated coupling between the electric-dipole ($\mathrm{ED}_y$) and magnetic-dipole ($\mathrm{MD}_x$) resonances. Their multipolar characteristics are analyzed in Supplementary Section S2 and Figures S2–S4. In addition, introducing the dielectric interconnections produces markedly different spectral shifts of the two resonances: the hybrid magnetic dipole ($\mathrm{hMD}_x$) mode undergoes a pronounced red-shift, whereas the hybrid electric dipole ($\mathrm{hED}_y$) mode is only weakly perturbed, resulting in a reversal of their spectral ordering. The measured transmission spectrum closely agrees with the simulated spectrum. Details of the measurements are provided in the Methods section. This distinction is further illustrated by the corresponding electric- and magnetic-field distributions of the constituent multipolar fields in Figures 2(b) and 2(c). The $MD_x$ resonance involves circulating displacement currents predominantly in the $yz$ plane and therefore interacts strongly with the $y$-oriented dielectric gratings. By contrast, the $hED_y$ mode exhibits comparatively weak spatial overlap with the interconnection regions. A detailed multipolar analysis of the Mie–grating metasurface is presented in Supplementary Section S3 and Figures S5–S7. Consequently, the dielectric interconnection regions between adjacent cylindrical resonators play a critical role in selectively tuning the constituent resonances and their hybridization, thereby providing a geometric degree of freedom for tailoring the collective electromagnetic response.

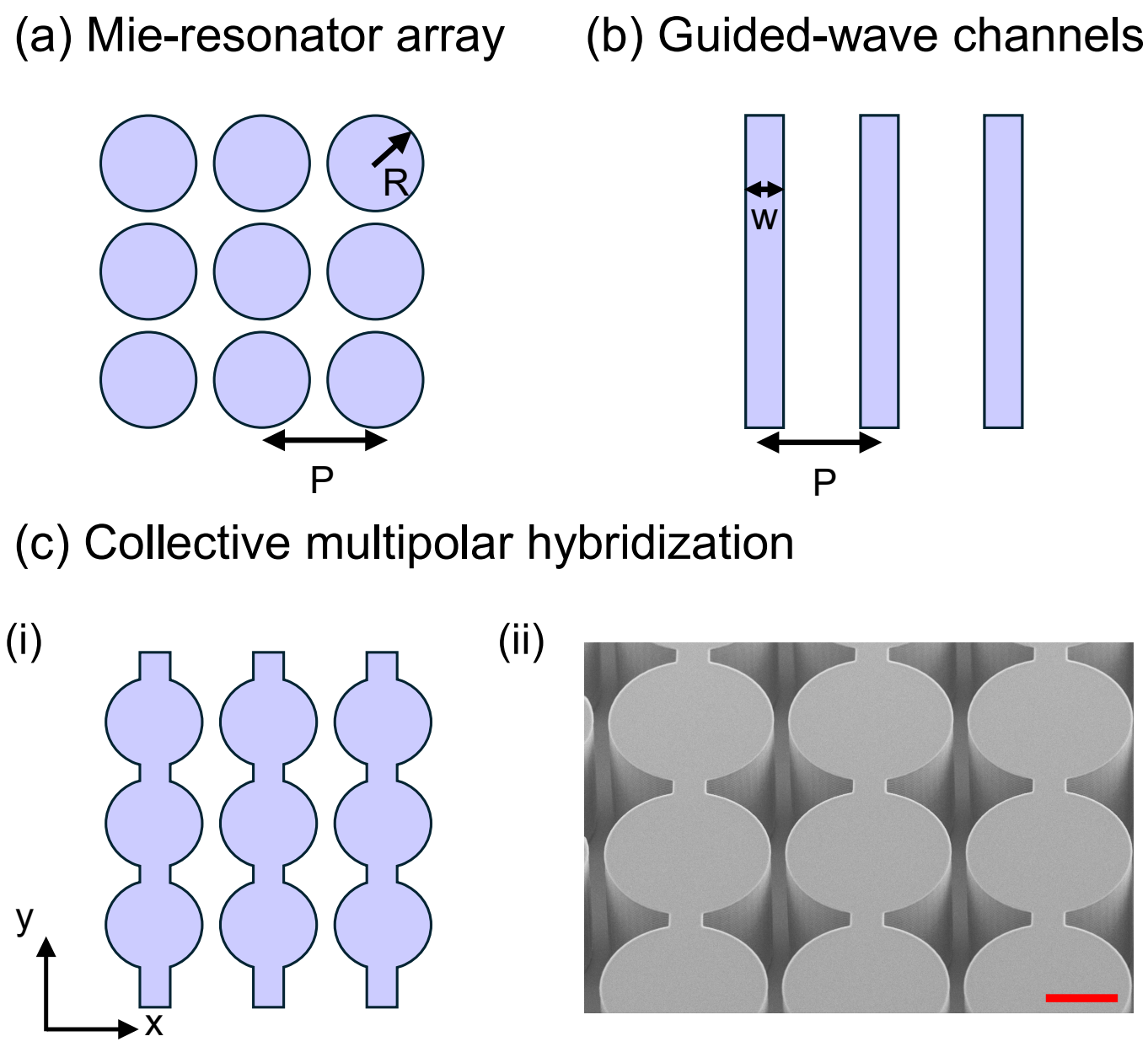

**FIGURE 1** | **Concept and realization of the free-standing THz Mie–grating metasurfaces.** (a) Schematic of a two-dimensional periodic array of cylindrical Mie resonators, where (R) and (P) denote the resonator radius and lattice period, respectively. (b) Schematic of a one-dimensional dielectric grating supporting guided-wave channels, characterized by the grating width (w) and lattice period (P). (c) Proposed hybrid Mie–grating metasurface integrating dielectric resonators and interconnecting bridges for engineering collective photonic modes and band dispersion. (i) Schematic of the free-standing Mie–grating metasurface, showing periodically arranged cylindrical resonators interconnected by dielectric gratings along the (y) direction. (ii) Scanning electron microscopy image of the fabricated metasurface. The scale bar represents 100$\mu$m.

To study how this mode-selective perturbation influences the photonic band dispersion of the hybrid Mie–grating, we performed angle-resolved transmission measurements and numerical simulations under TE- and TM-polarized excitation along two orthogonal momentum directions. Details of measurements and simulations can be found in Methods. The measured dispersion maps closely reproduce the principal features of the simulated band structures in Figures 2(d) and 2(e). Under TE-polarized incidence in the $xz$ plane, as illustrated in Figure 2(d,i), varying the incidence angle $\theta_x$ changes the in-plane Bloch momentum $k_x$ while maintaining the electric field along the $y$ direction. The $hMD_x$ branch exhibits remarkably weak frequency dispersion, arising from the compensation of momentum-dependent dispersive contributions through collective multipolar hybridization, thereby forming a quasi-flat photonic band. In contrast, the $hED_y$ branch retains pronounced dispersion, as shown in Figure 2(d,ii,iii).

A complementary band evolution emerges under TM-polarized incidence in the $yz$ plane, where the incidence angle $\theta_y$ determines the in-plane Bloch momentum $k_y$ displayed in Figure 2(e). In contrast to the TE configuration, the $hED_y$ branch becomes nearly dispersionless across the entire $k_y$ momentum range, whereas the $hMD_x$ branch exhibits pronounced linear dispersion, as shown in Figure 2(e,ii,iii). Notably, the grating-induced spectral shift of the $hMD_x$ branch brings it into close spectral proximity with the electric-quadrupole ($EQ_z$) branch, which supports a symmetry-protected bound state in the continuum (BIC) at the Γ point. From a symmetry-based perspective, the $EQ_z$ mode is decoupled from the available free-space radiation channels at the Γ point owing to their incompatible symmetries. From a collective electromagnetic perspective, this radiation suppression can be understood in terms of destructive interference among the radiation amplitudes associated with different multipolar contributions, leading to a dark mode [57, 27]. Away from the Γ point, this symmetry protection is lifted, allowing the $EQ_z$ mode to couple to radiative channels and emerge as a dispersive resonance. The characteristic $E_z$ field distribution of the $EQ_z$ mode at finite $k_y$ is illustrated in the inset of Figure 2(e,ii), with its multipolar composition further analyzed in Supplementary Figure S8. These results highlight how collective multipolar interactions govern both the dispersive and radiative properties of photonic modes within the same hybrid Mie–grating structure.

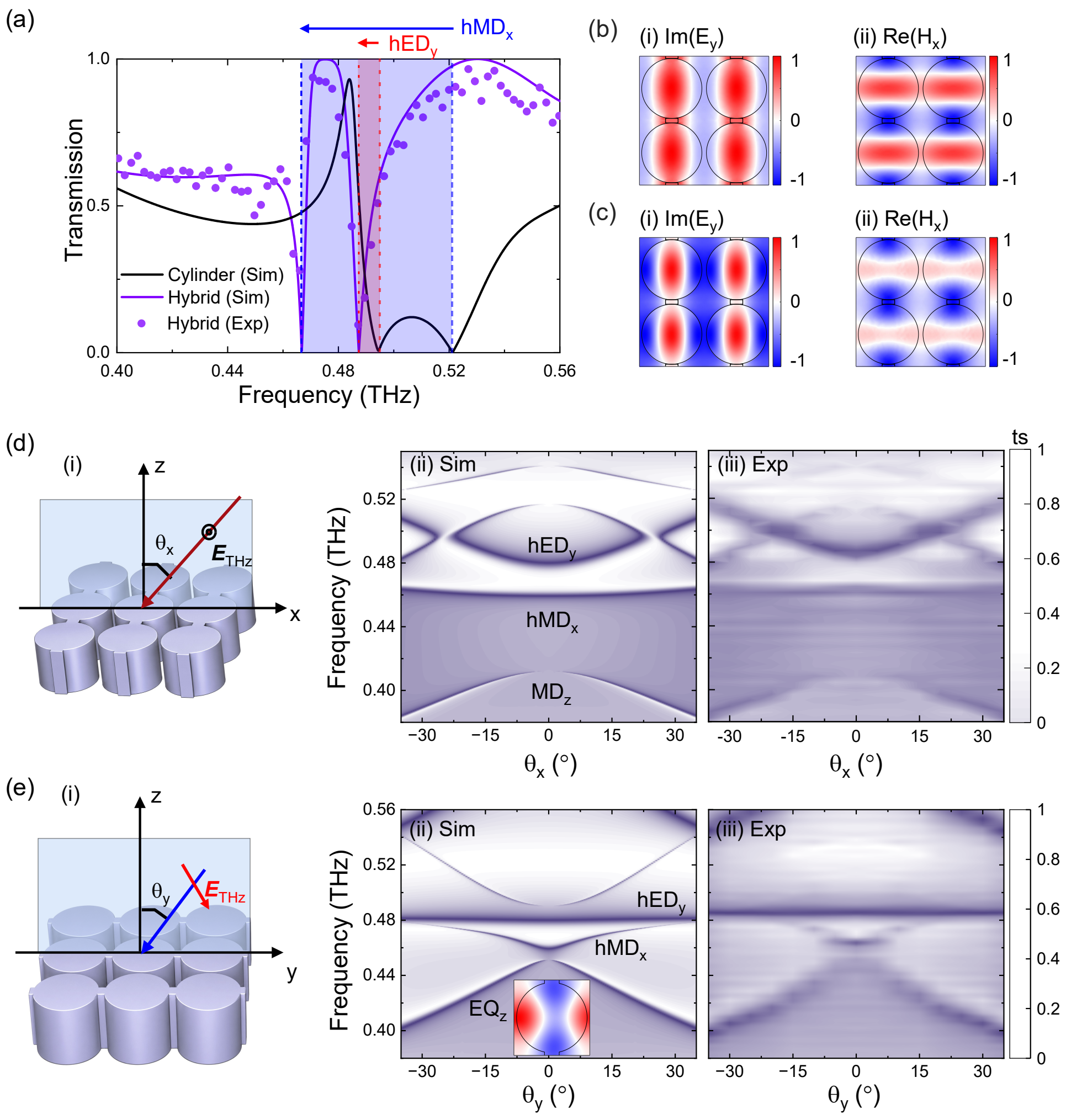

**FIGURE 2** | **Anisotropic quasi-flat photonic bands in a hybrid Mie–grating metasurface.** (a) Simulated and measured transmission spectra of the metasurface with the grating and simulated spectrum of the metasurface without the grating, revealing two hybrid eigenmodes. These eigenmodes arise from the lattice-mediated hybridization between the in-plane magnetic-dipole ($\mathrm{MD_x}$) and electric-dipole ($\mathrm{ED_y}$) resonances: the $\mathrm{hMD_x}$ mode, with stronger $\mathrm{MD_x}$ character, and the $\mathrm{hED_y}$ mode, with stronger $\mathrm{ED_y}$ character. The shaded regions highlight the grating-induced resonance shifts and reordering of the two hybrid modes. (b,c) Simulated electric- and magnetic-field distributions, $\mathrm{Im}(E_y)$ **(i)** and $\mathrm{Re}(H_x)$ **(ii)**, evaluated at half the resonator height for the $\mathrm{hMD}_x$ **(b)** and $\mathrm{hED}_y$ **(c)** modes, respectively. The distinct field distributions reveal different degrees of spatial overlap with the connecting grating, enabling selective tuning of the hybrid resonances. (d) Angle-resolved response under TE-polarized incidence for rotation about the $y$ axis. (i) Schematic of the measurement geometry, with the incident THz wave tilted by an angle $\theta_x$. (ii) Simulated angle-resolved transmission map showing the dispersions of the $\mathrm{hMD_x}$, $\mathrm{hED_y}$, and lowest-order magnetic-dipole ($\mathrm{MD_z}$) bands. The $\mathrm{hMD_x}$ band exhibits quasi-flat dispersion. (iii) Experimental angle-resolved transmission map confirming the quasi-flat $\mathrm{hMD_x}$ band. (e) Angle-resolved response under TM-polarized incidence for rotation about the $x$ axis. (i) Schematic of the measurement geometry, with the incident THz wave tilted by an angle $\theta_y$. (ii) Simulated angle-resolved transmission map showing that grating-induced closure of the spectral gap between the $\mathrm{hMD_x}$ and electric-quadrupole ($\mathrm{EQ_z}$) bands produces a Dirac-like cone. A second quasi-flat dispersion associated with the $\mathrm{hED_y}$ hybrid mode is also observed. The inset shows the simulated $E_z$ field distribution of the $\mathrm{EQ_z}$ mode at $\theta_y = 2°$. (iii) Experimental angle-resolved transmission map confirming the Dirac-like cone of the $\mathrm{hMD_x}$ band and the quasi-flat dispersion associated with the $\mathrm{hED_y}$ hybrid mode.

## 2.2 Visualization of LDOS enabled by flatbands

The combination of weak dispersion and narrow linewidth of the quasi-flatband $\mathrm{hMD}_x$ resonance is expected to enhance the local density of optical states experienced by an appropriately positioned and oriented emitter [58, 59, 60]. Access to the partial-LDOS (PLDOS) provides insight into the density of available photonic states at a given frequency, position, and polarization, revealing how the resonant modes shape the local electromagnetic environment and their potential for enhanced light–matter interactions [61, 62, 63]. To experimentally investigate the PLDOS in the THz Mie–grating metasurface, we employed phase-resolved

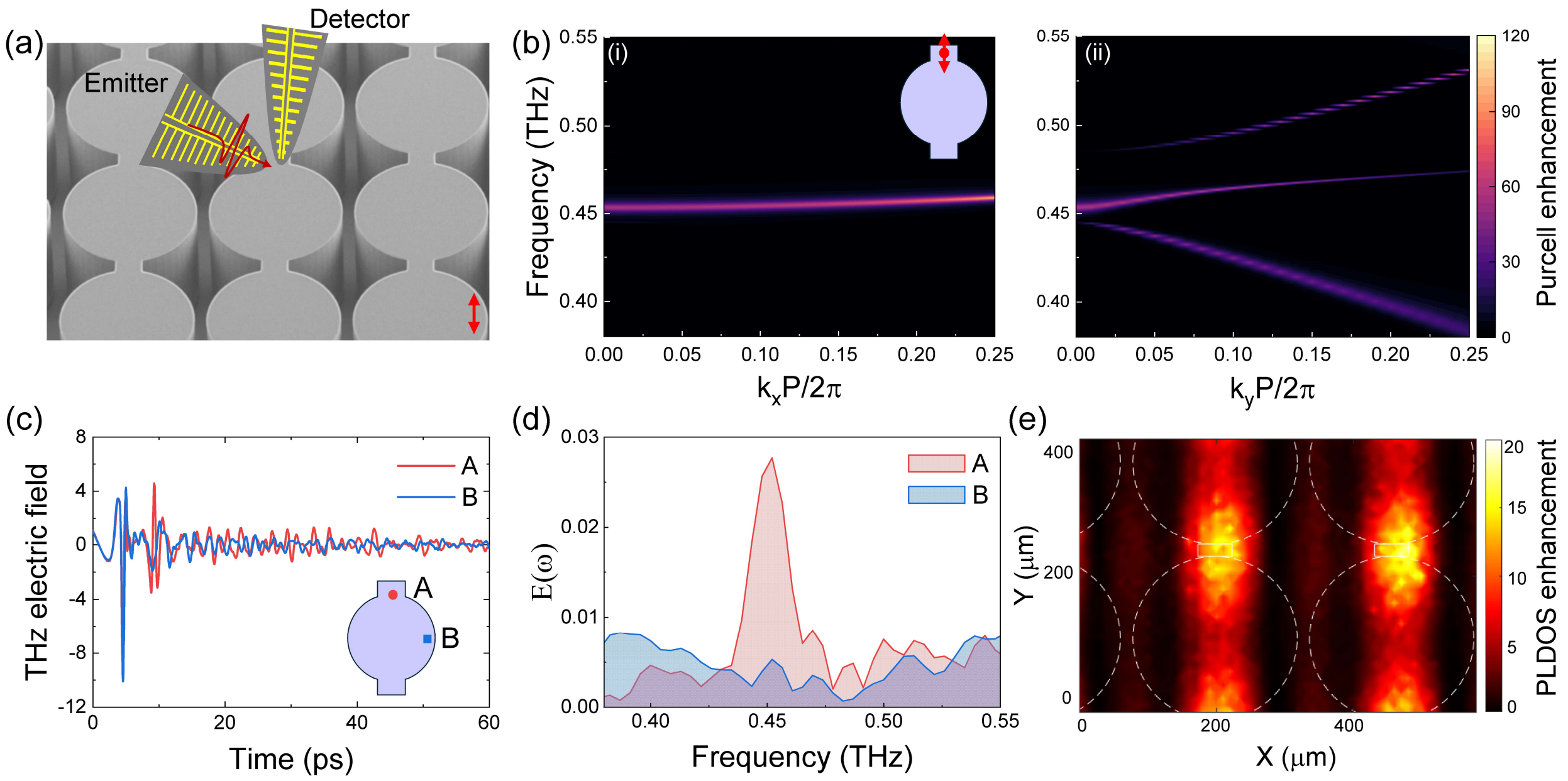

**FIGURE 3** | **Enhanced partial local density of optical states associated with the quasi-flat** $hMD_x$ **band.** (a) Schematic illustration of the THz near-field double-dipole measurements performed on the free-standing THz Mie–grating metasurface. Two $y$-oriented electric-dipole probes with subwavelength separation and positioned above the metasurface provide localized excitation and detection, respectively. (b) Calculated Purcell enhancement for a $y$-oriented electric dipole located above the center of the dielectric interconnection region, resolved as a function of frequency and normalized in-plane Bloch momentum along (i) $k_xP/2\pi$ and (ii) $k_yP/2\pi$. A pronounced enhancement follows the quasi-flat $hMD_x$ band, indicating the coupling between the localized emitter and the collective photonic mode. (c) THz transients of the time-dependent electric field measured with the microprobes respectively locating at two representative positions, A and B, as indicated in the inset. (d) Corresponding Fourier-amplitude spectra of the transients in (c), showing a pronounced resonant enhancement near 0.45 THz at position A, consistent with the quasi-flat $hMD_x$ resonance. (e) Experimentally reconstructed spatial map of the PLDOS enhancement at the $hMD_x$ resonance. The enhanced response extends along the dielectric gratings, revealing spatially distributed electromagnetic interactions associated with the collective mode. Dashed contours indicate the underlying resonator geometry.

THz near-field spectroscopy using two subwavelength electric-dipole probes, enabling direct measurements of the PLDOS in different photonic environments [64, 65]. A schematic of the experimental setup is shown in Figure 3(a). The two probes, serving as the excitation source and detector, respectively, are oriented along the $y$-direction and laterally separated by 10 $\mu$m to minimize optical crosstalk from the laser used to generate and detect THz radiation in the probes (see Methods section). Both probes are positioned at a height of 10 $\mu$m above the metasurface. This deeply subwavelength configuration enables the localized excitation and probing of the resonant electromagnetic response. Following our previously established dual-probe methodology for the direct measurement of the PLDOS [65], the PLDOS enhancement is experimentally estimated from the normalized imaginary component of the complex electric field, $\mathrm{Im}[E_y(\mathbf{r},\omega)]/\mathrm{Im}[E_{y,0}(\mathbf{r},\omega)]$, where $E_{y,0}$ denotes the reference field measured in free space under identical probe conditions. Further experimental details are provided in the Methods section. The corresponding PLDOS characterizes the electromagnetic environment experienced by the dipole and depends sensitively on its position and orientation. For a $y$-oriented electric dipole located at $\mathbf{r}_0$, the PLDOS can be expressed in terms of the diagonal term of the dyadic Green's function along the y-direction [66],

$$\rho_y(\mathbf{r}_0,\omega) = \frac{2\omega}{\pi c^2}\,\mathrm{Im}\,G_{yy}(\mathbf{r}_0,\mathbf{r}_0;\omega). \tag{1}$$

Unlike far-field plane-wave illumination, which selectively excites Bloch modes at a well-defined in-plane momentum, the localized electric dipole can couple to collective modes spanning a broad range of wavevectors [67]. For a two-dimensional periodic metasurface, the local electromagnetic response can therefore be decomposed into contributions from different in-plane Bloch wavevectors [68, 69],

$$\rho_y(\mathbf{r}_0,\omega) = \int_{\mathrm{BZ}} \frac{d^2\mathbf{k}_\parallel}{(2\pi)^2}\rho_y(\mathbf{r}_0,\mathbf{k}_\parallel,\omega), \tag{2}$$

where $\rho_y(\mathbf{r}_0,\mathbf{k}_\parallel,\omega)$ represents the wavevector-resolved contribution to the LDOS, defined with a consistent Bloch normalization. Since the Purcell enhancement is directly proportional to the PLDOS [70, 71], we calculated the Bloch-wavevector-resolved Purcell enhancement to investigate the momentum-space origin of the enhanced PLDOS, with particular emphasis on the contribution of the dielectric interconnections defining the grating. A $y$-oriented electric dipole was positioned above the center of a dielectric interconnection, and the Purcell enhancement was calculated along two orthogonal momentum directions by varying $k_x$ at $k_y = 0$ and $k_y$ at $k_x = 0$ as shown in Figures 3(b,i) and 3(b,ii), respectively. Along $k_x$, a pronounced enhancement occurs near 0.45 THz and extends over a broad momentum range, closely following the quasi-flat $hMD_x$ band displayed in Figure 3(b,i). In contrast, the response along $k_y$ follows strongly dispersive resonant branches in Figure 3(b,ii). The agreement with the band dispersions identified by far-field spectroscopy confirms that localized electric-dipole excitation can access these collective Bloch resonances. Notably, despite its predominantly magnetic-dipolar character, the quasi-flat $hMD_x$ mode couples to the electric dipole through its finite local $E_y$ component. This combination of weak dispersion and nonzero electric-dipole overlap identifies the quasi-flatband as a potential source of enhanced PLDOS at the interconnection region.

To experimentally investigate the spatial dependence of the near-field local electromagnetic response with the dipole excitation, Figure 3(c) shows time-domain THz electric-field transients at two representative positions (A and B) on the hybrid Mie–grating. A and B are located near the dielectric interconnection and the lateral edge of the cylindrical resonator, respectively, as illustrated in the inset of Figure 3(c). At position A, the measured transient exhibits pronounced oscillations persisting beyond the initial excitation, whereas the response at position B decays more rapidly. The corresponding frequency-domain amplitude spectra in Figure 3(d) reveal a pronounced resonance near 0.45 THz at position A, consistent with the quasi-flat-band mode identified in the TE dispersion. Compared with the far-field measurement, this resonance exhibits a slight frequency shift, potentially arising from differences in the excited wavevector components and perturbations introduced by the nearby dipole probes. Furthermore, the pronounced contrast between positions A and B highlights the spatially nonuniform electric-field response associated with the quasi-flat mode.

We further raster-scanned the near-field probe assembly across multiple unit cells of the hybrid Mie–grating metasurface with a step size of 10 $\mu$m to spatially resolve the PLDOS associated with the quasi-flat $hMD_x$ mode. Figure 3(e) shows the spatial distribution of PLDOS enhancement extracted at the frequency of the quasi-flatband. It is clear that the strongest enhancement is observed along the dielectric interconnections between neighboring cylindrical resonators. Although the quasi-flatband is dominated by the $hMD_x$ mode, it can still be excited by a $y$-polarized electric dipole, highlighting the crucial role of the local $E_y$ field component in mediating the coupling between the dipole and the collective quasi-flat-band mode. This coupling further underscores the importance of the grating geometry, particularly the dielectric interconnection region, in shaping the local electric-field distribution and facilitating the hybridization between the electric dipole and the predominantly magnetic collective mode. Furthermore, the PLDOS enhancement extends spatially along the grating, suggesting that the electromagnetic response is distributed across the interconnected resonators rather than being confined to individual cylinders, consistent with the collective nature of the quasi-flat-band resonance.

## 2.3 Grating-engineered high-$Q$ flatbands

Having established the critical role of the dielectric interconnection region in shaping the multipolar hybridization underlying the quasi-flatband, we next investigated how varying the grating width tunes the spectral and radiative properties of the collective resonance by modifying the spatial overlap of its constituent multipolar field components while keeping the lattice constant fixed. To systematically examine this geometric dependence, we considered three other representative grating widths, w = 25, 75, and 100 $\mu$m, and characterized their angle-resolved transmission spectra under both TE- and TM-polarized incidence, as shown in Figures 4(a) and 4(b). Optical images of the fabricated structures are provided in Supplementary Figure S9, and the corresponding simulated dispersion maps are provided in Supplementary Section S4 and Figure S10. When increasing the grating width, the quasi-flat $hMD_x$ band exhibits a pronounced red-shift while preserving its weakly dispersive character under TE excitation. This spectral shift alters the frequency separation $\Delta f$ between the quasi-flat $hMD_x$ band and the dispersive electric-quadrupole ($EQ_z$) branch observed under TM-polarized excitation, thereby enabling geometric control over their spectral proximity and the resulting Dirac-like band feature.

The mode-selective spectral evolution is further confirmed by the measured and simulated normal-incidence transmission spectra in Figure 4(c) and 4(d), respectively. With increasing grating width w, the $hMD_x$ resonance exhibits a pronounced red-shift accompanied by progressive linewidth narrowing, indicating a strong dependence of its spectral and radiative properties on the dielectric interconnection region. This sensitivity can be attributed to the enhanced spatial overlap and modified coupling between its constituent electric- and magnetic-dipole field components as the interconnection width increases [72, 73, 74]. In contrast, the $hED_y$ resonance remains comparatively insensitive to variations in w. Although the $hED_y$ resonance also exhibits hybrid electric- and magnetic-dipole characteristics, its electromagnetic response is predominantly governed by the individual cylindrical resonators, whose geometry remains unchanged. These contrasting behaviors indicate that varying the grating width selectively modifies the multipolar hybridization rather than merely producing a uniform change in the effective dielectric environment.

Beyond spectral tuning, the dielectric interconnections of the hybrid Mie–grating structure also provide substantial control over the resonance decay rates. Figure 4(e) compares the simulated and measured $Q$ factors of the two hybrid modes. While the $Q$ factor of $hED_y$ remains nearly unchanged, that of $hMD_x$ increases markedly with grating width, approaching 500 in simulation when the grating width is 110 $\mu$m. This enhancement is consistent with reduced radiative leakage resulting from changes in the coupling of the hybrid magnetic-dipolar mode to outgoing radiation channels. The lower experimentally observed $Q$ factors may be attributed to intrinsic material absorption and fabrication imperfections, which introduce additional dissipative and scattering losses [75, 76]. Furthermore, finite-size effects in the array induce boundary-related radiative leakage, further reducing the achievable $Q$ factors compared with those predicted for ideal infinite periodic structures [77, 78]. Nevertheless, the consistent trends demonstrate mode-selective control over the resonance linewidth through the dielectric interconnection.

## 2.4 Purcell enhancement: Interplay between modal lifetime and mode volume

The substantial enhancement of the modal $Q$ factor of the quasi-flatband motivates further investigation of how its increased resonance lifetime influences the local electric-dipole response. To investigate this relationship, we measured the PLDOS enhancement spectra of the Mie–grating metasurfaces with three representative grating widths, w = 25, 75, and 100 $\mu$m, and compared them with the corresponding calculated Purcell enhancement spectra, as shown in Figure 5(a). Both responses exhibit a progressive red-shift and spectral narrowing of the hMD$_x$ resonance with increasing grating width. Interestingly, the PLDOS and Purcell enhancements exhibit opposite trends in magnitude: the measured near-field enhancement progressively decreases, whereas the calculated Purcell enhancement increases as the resonance linewidth narrows. This contrasting behavior arises from the distinct electromagnetic responses probed by the two approaches. The near-field experiment captures the transient source-to-detector response of a finite metasurface, which is influenced by excitation and detection efficiencies and the finite temporal acquisition window [79, 80]. In contrast, the calculated Purcell enhancement describes the steady-state response of a $y$-oriented electric dipole per unit cell under Γ-point Bloch-periodic boundary conditions [81, 82]. Consequently, the measured near-field enhancement and calculated Purcell enhancement may exhibit contrasting trends in magnitude despite originating from the same underlying collective resonance and exhibiting similar spectral evolution.

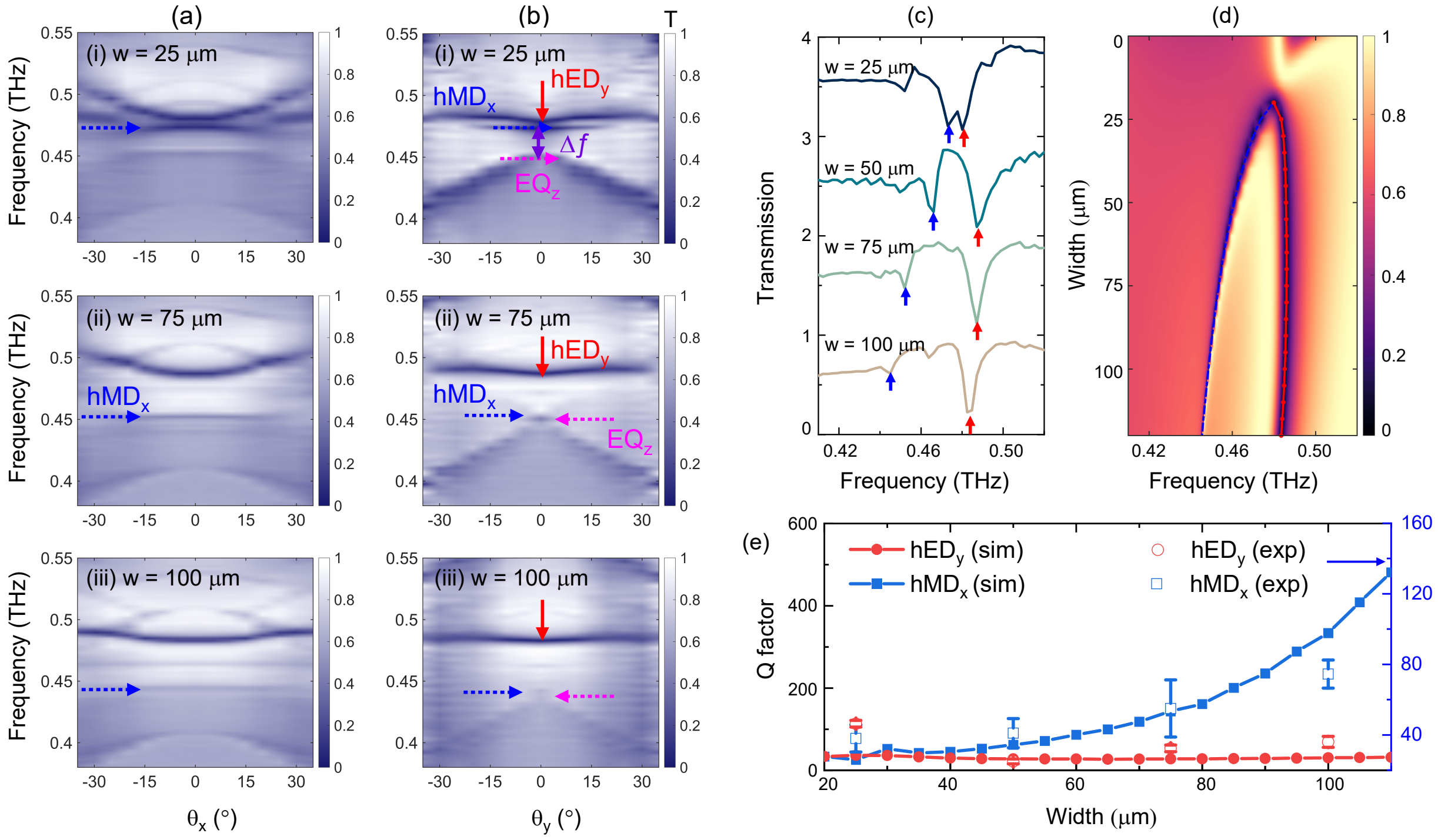

**FIGURE 4** | **Grating-width-tuned evolution of hybrid photonic bands and *Q* factors.** (a,b) Experimentally measured angle-resolved transmission spectra under (a) TE- and (b) TM-polarized incidence for grating widths of (i) w = 25 $\mu$m, (ii) w = 75 $\mu$m, and (iii) w = 100 $\mu$m. Under TE excitation, increasing the grating width progressively red-shifts and flattens the hMD$_x$ band. Under TM excitation, the grating-induced red-shift of hMD$_x$ reduces its spectral separation $\Delta f$ from the comparatively weakly perturbed EQ$_z$ branch. (c) Experimentally measured normal-incidence transmission spectra for different grating widths, vertically offset for clarity. Blue and red arrows identify the hMD$_x$ and hED$_y$ resonances, respectively, highlighting their contrasting spectral shifts. (d) Simulated normal-incidence transmission as a function of frequency and grating width, illustrating the continuous evolution and spectral reordering of the hybrid resonances. (e) Simulated (filled symbols) and experimentally extracted (open symbols) $Q$ factors of the hMD$_x$ (blue) and hED$_y$ (red) resonances as functions of grating width. The left and right axes correspond to simulated and experimental $Q$ factors, respectively. The hMD$_x$ $Q$ factor increases markedly with grating width, whereas that of hED$_y$ remains comparatively unchanged.

To further quantify the influence of grating geometry on the intrinsic light–matter interaction, the grating-width-dependent evolution of the calculated Purcell enhancement is presented in Figure 5(b). Increasing the grating width w induces a continuous red-shift and progressive spectral narrowing of the quasi-flat hMD$_x$ resonance, consistent with the experimental observations presented in Figures 4(c)- 4(e). Notably, this spectral evolution is accompanied by a pronounced increase in the peak Purcell enhancement, indicating stronger resonant light–matter interactions as the linewidth narrows. Figure 5(c) further examines this relationship by plotting the peak Purcell enhancement against the $Q$ factor of the quasi-flat hMD$_x$ resonance for different grating widths. Although the enhancement increases with $Q$, its growth gradually slows, exhibiting a tendency toward saturation at higher $Q$ factors. Within the conventional single-mode cavity approximation, the peak Purcell enhancement scales as $F_P^{peak} \propto Q/V_{eff}$, where $V_{eff}$ denotes the effective mode volume experienced by the emitter [83, 84]. As the grating width increases, the hybrid electromagnetic mode extends over a larger spatial region, leading to an increase in $V_{eff}$ and a corresponding reduction in the coupling strength between the single dipole emitter and the cavity mode. Consequently, the observed tendency toward saturation of the Purcell enhancement suggests that the increase in mode volume partially offsets the enhancement associated with the increasing $Q$ factor. This behavior highlights the interplay between modal confinement and resonance lifetime in governing the local light–matter interaction.

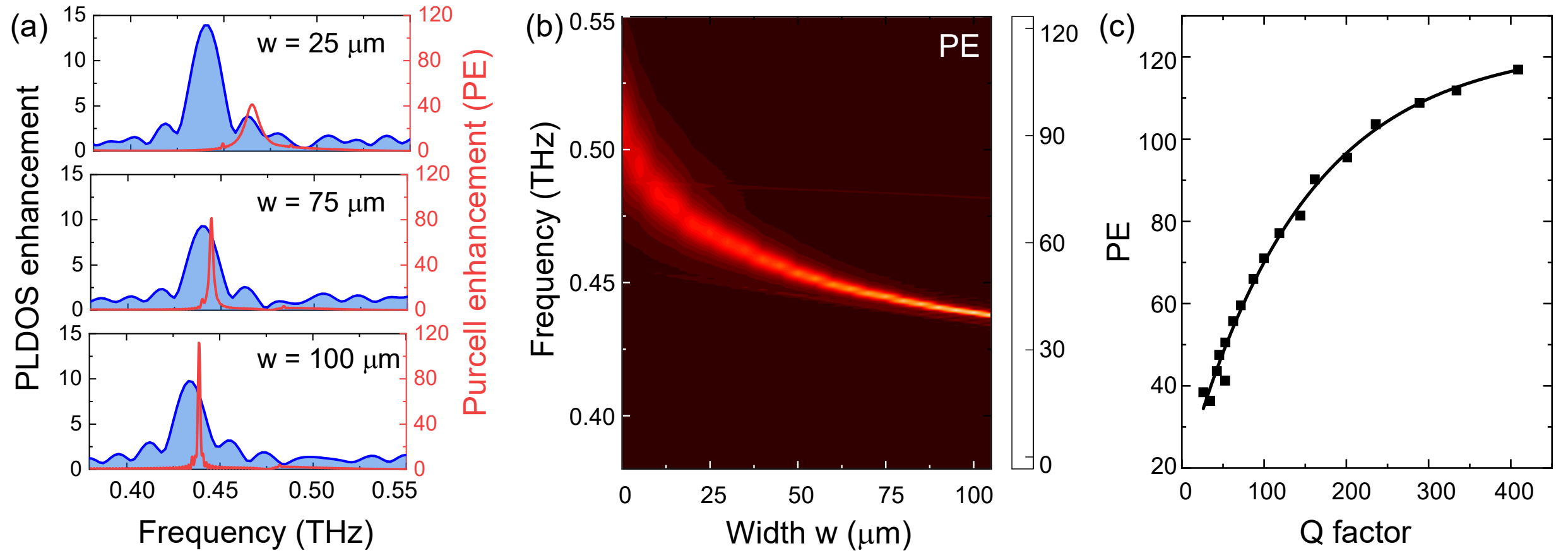

**FIGURE 5** | **Grating-width-controlled radiative confinement and electric-dipole Purcell enhancement of the** $hMD_x$ **resonance.** (a) PLDOS enhancement spectra obtained from THz near-field microscopy with a finite time-delay scan range (blue, left axis) and calculated steady-state Purcell enhancement spectra (red, right axis) for grating widths of w = 25, 75, and 100 $\mu$m. The calculations consider a $y$-oriented electric dipole positioned above the center of a dielectric grating under Γ-point Bloch-periodic conditions. (b) Calculated Purcell enhancement as a function of frequency and grating width, illustrating the progressive red-shift, spectral narrowing, and increasing peak enhancement of the $hMD_x$ resonance. (c) Peak Γ-point Purcell enhancement as a function of the $Q$ factor of the $hMD_x$ resonance for different grating widths. The enhancement initially increases with $Q$ but exhibits a tendency toward saturation at higher $Q$ factors, highlighting the interplay between radiative confinement and local emitter–mode coupling.

Together, these results demonstrate that dielectric interconnections provide an effective geometric degree of freedom for enhancing the $Q$ factor of quasi-flat collective photonic modes while tailoring their coupling to localized electric dipoles in hybrid Mie–grating metasurfaces. The interplay among weak band dispersion, prolonged modal lifetime, and spatial emitter–mode overlap highlights the importance of jointly engineering momentum-space band characteristics and local electromagnetic field distributions to control light–matter interactions in nonlocal photonic metasurfaces.

## 3 Conclusions

We have designed and experimentally demonstrated a free-standing THz Mie–grating metasurface in which dielectric interconnections between Mie-resonators act as active geometric elements for engineering collective quasi-flat photonic bands through multipolar hybridization. By selectively perturbing electric- and magnetic-dipolar resonances, the dielectric interconnections support strongly anisotropic band structures featuring quasi-flat modes along distinct momentum directions and a Dirac-like spectral feature associated with the hybrid magnetic-dipolar and electric-quadrupolar branches. THz near-field measurements reveal long-lived resonant oscillations and spatially-resolved PLDOS enhancement, with the strongest response concentrated along the interconnections. These observations establish the local electric-dipole accessibility of the collective quasi-flat magnetic-dipolar mode despite its predominantly magnetic character.

Systematic variation of the grating width enables continuous spectral tuning and substantial enhancement of the modal $Q$ factor of the quasi-flat $hMD_x$ band, while preserving its weakly dispersive character. Complementary Bloch-resolved Purcell calculations further reveal that the local resonant electric-dipole response increases with $Q$ factor but exhibits a tendency toward saturation, highlighting the interplay between modal lifetime and effective mode volume. Together, these findings establish engineering the dielectric interconnection regions of the Mie–grating metasurface as an effective strategy for jointly tailoring the spectral, modal, and local electromagnetic properties of high-$Q$ quasi-flat photonic bands. Beyond the THz regime, this approach provides a promising framework for designing nonlocal dielectric metasurfaces with tailored collective resonances and localized light–matter interactions.

## 4 Methods

### 4.1 Sample fabrication

High-resistivity silicon wafers (resistivity $\geq$ 5000 $\Omega \cdot$ cm), fabricated via the float-zone method, were first subjected to a standard RCA cleaning process to remove organic, ionic, and metallic contaminants. A ~1050 nm-thick $SiO_2$ layer was subsequently grown by wet thermal oxidation. Metasurface patterns were defined using laser direct-write photolithography. A 2 $\mu$m-thick layer of AZ 5214E photoresist was spin-coated onto the wafer, exposed with a 375 nm laser direct-writing system (MLA 150, Heidelberg), and developed in AZ 726 MIF. The $SiO_2$ layer was then etched via reactive ion etching (RIE; SPTS Pegasus), yielding anisotropic profiles. After removal of the residual photoresist, the patterned $SiO_2$ hard mask was used to transfer the metasurface patterns into the silicon substrate through deep reactive ion etching (DRIE) using the Bosch process. This process, based on alternating passivation and etching cycles, enabled the formation of high-aspect-ratio structures with steep sidewalls. Lastly, the remaining $SiO_2$ mask was removed using buffered hydrofluoric acid, followed by rinsing in deionized water.

### 4.2 Far-field simulations of plane wave excitation

The optical response of the hybrid Mie–grating metasurface was numerically investigated using the frequency-domain solver of CST Studio Suite, based on the finite element method (FEM). A single unit cell was modeled with periodic boundary conditions in the in-plane directions and plane-wave excitation at normal and oblique incidence. Transmittance spectra and angle-resolved dispersion were obtained by sweeping the incidence angle, while the corresponding electric-field distributions were calculated to characterize the resonant modes. The free-standing silicon metasurface was modeled as a lossless dielectric with a relative permittivity of $\varepsilon_{\mathrm{Si}} = 11.7$, surrounded by a homogeneous air environment with a refractive index of $n = 1$.

### 4.3 Dependence of THz far-field measurements on the incidence angle

The fabricated free-standing silicon metasurfaces were characterized in transmission using a THz time-domain spectrometer (Menlo K15). Broadband THz pulses were generated and detected using photoconductive antennas driven by femtosecond laser pulses at a wavelength of $\lambda = 1.5\ \mu$m. Four polyethylene lenses were employed to collimate and focus the THz beam onto the sample, resulting in a spot size of approximately 5 mm. Two wire-grid THz polarizers mounted on rotation stages were positioned before and after the sample to define and analyze the linear polarization of the transmitted radiation. Angle-resolved TE- and TM-polarized transmission spectra were obtained by rotating the sample about the $y$ and $x$ axes shown in Figure 2, respectively, while keeping the incident THz polarization fixed. The incidence angle was defined as the angle between the incident THz beam and the surface normal of the metasurface.

### 4.4 Measurements of spatially resolved LDOS using THz near-field microscopy

The local density of optical states was investigated using a THz near-field microscopy setup comprising two commercially available photoconductive microprobes (TeraSpike TD-800-X-HR-WT, Protemics GmbH), serving as a localized emitter and detector, respectively. Each microprobe consists of a low-temperature-grown GaAs (LT-GaAs) photoconductive element integrated with microstructured electrodes. Broadband THz pulses were generated by illuminating the biased emitter with femtosecond laser pulses (100 fs duration, 780 nm central wavelength), while the unbiased detector probe was used to measure the THz electric field. The localized THz emitter had an effective excitation size of approximately 20 $\mu$m, substantially smaller than both the metasurface periodicity (258 $\mu$m) and the free-space wavelength of the quasi-flat-band resonance ($\lambda_0 \approx 650\ \mu$m). The localized excitation could therefore be approximated as an electric point dipole. By positioning the emitter and detector in close proximity, with a separation much smaller than the THz wavelength, the electric field generated by the dipole could be probed near its source position. This configuration provides access to the partial local density of optical states (PLDOS) associated with the dipole orientation, which is related to the imaginary component of the local electric field at the emitter position.

The time-resolved electric field $E(t)$ was recorded by scanning an optical delay stage, enabling phase-sensitive detection of the THz transient signal. The corresponding complex frequency-domain field $E(\omega)$ was obtained by Fourier transformation. A temporal acquisition window of approximately 80 ps provided a nominal frequency resolution of 12.5 GHz. Zero padding was applied prior to Fourier transformation to increase the frequency-domain sampling density without changing the intrinsic spectral resolution.

Spatially resolved measurements were performed by raster scanning the metasurface with a step size of 10 $\mu$m while recording the THz transient signal at each position. Fourier transformation of the time-domain signals yielded spatially resolved complex electric-field spectra, containing both amplitude and phase information.

### 4.5 Dipole excitation and Purcell enhancement

Numerical simulations of the localized electric-dipole response were performed using the finite-difference time-domain method implemented in Ansys Lumerical FDTD. A single unit cell of the hybrid Mie-grating metasurface was modeled with Bloch-periodic boundary conditions along the $x$ and $y$ directions and perfectly matched layers (PMLs) along the out-of-plane $z$ direction. A point electric-dipole source polarized along the $y$ direction was positioned above the center of the dielectric grating to excite the hybrid collective modes.

To characterize the momentum-dependent dipole response, the in-plane Bloch wavevector was independently varied along the $k_x$ and $k_y$ directions, with the orthogonal wavevector component fixed at zero. For the grating-width-dependent calculations, both in-plane wavevector components were fixed at zero, corresponding to the $\Gamma$ point.

The frequency-dependent dipole response was evaluated from the electromagnetic fields and source power to determine the Purcell enhancement relative to a reference environment. Under Bloch-periodic boundary conditions, the dipole excitation represents a phase-coherent periodic array of emitters rather than an isolated point dipole. Consequently, the calculated enhancement describes the Bloch-wavevector-resolved dipole response, whereas obtaining the total Purcell enhancement of an isolated emitter requires integration over the two-dimensional Brillouin zone.

## Author contributions

J.J. and T.X.H contributed equally to this work. T.X.H., Y.K. and J.G.R. conceived the original idea and initiated the project. S.Z. and J.J. carried out sample fabrication. J.J. designed and constructed the experimental setup and performed THz measurements. J.J., T.X.H., A.N., and K.W. performed the theoretical calculations and numerical simulations. J.J. and T.X.H. analyzed the results and discussed the underlying physics and potential applications with all co-authors. J.J. and T.X.H. wrote the original manuscript, while J.G.R. and Y.K. reviewed and revised it. All authors discussed the results and approved the final manuscript.

## Acknowledgments

This project has received funding from the PhotonDelta National Growth Fund program. K.W. acknowledges financial support from the European Union through the SCOLED project (Grant Agreement No. 101098813). A.N. acknowledges support from the National Research Foundation, Singapore, and A*STAR under its Frontier Competitive Research Programme (Grant No. NRF-F-CRP-2024-0090).

## Conflicts of Interest

The authors declare no conflicts of interest.

## Data Availability Statement

The data that support the findings of this study are available from the corresponding authors upon request.

## References

1. X. Huang, Y. Lai, Z. H. Hang, H. Zheng, and C. T. Chan. "Dirac cones induced by accidental degeneracy in photonic crystals and zero-refractive-index materials." *Nat Mater* 10, no. 8 (2011): 582–586.
2. H. S. Nguyen, F. Dubois, T. Deschamps, S. Cueff, A. Pardon, J. L. Leclercq, C. Seassal, X. Letartre, and P. Viktorovitch. "Symmetry breaking in photonic crystals: on-demand dispersion from flatband to Dirac cones." *Phys Rev Lett* 120, no. 6 (2018): 066102.
3. C. Danieli, A. Andreanov, D. Leykam, and S. Flach. "Flat band fine-tuning and its photonic applications." *Nanophotonics* 13, no. 21 (2024): 3925–3944.
4. R. Heilmann, J. Lehikoinen, S. Eyvazi, E. A. Mamonov, and P. Törmä. "Chains of nanoparticles for flat-band emission and lasing." *Nano Lett* 26, no. 17 (2026): 5960–5966.
5. I. Amelio and N. Goldman. "Lasing in Non-Hermitian flat bands: quantum geometry, coherence, and the fate of Kardar-Parisi-Zhang physics." *Phys Rev Lett* 132, no. 18 (2024): 186902.
6. F. Jin, J. Ren, S. Zanotti, P. Cui, H. Zheng, J. Liang, J. Ni, Z. Wang, F. Shi, C. W. Qiu, G. Chang, T. C. H. Liew, and R. Su. "Exciton polariton condensation in a perovskite moiré flat band at room temperature." *Sci Adv* 11, no. 32 (2025): eadx2361.
7. S. Mukherjee, A. Spracklen, D. Choudhury, N. Goldman, P. Öhberg, E. Andersson, and R. R. Thomson. "Observation of a localized flat-band state in a photonic Lieb lattice." *Phys Rev Lett* 114, no. 24 (2015): 245504.
8. Y. Yang, C. Roques-Carmes, S. E. Kooi, H. Tang, J. Beroz, E. Mazur, I. Kaminer, J. D. Joannopoulos, and M. Soljačić. "Photonic flat-band resonances for free-electron radiation." *Nature* 613, no. 7942 (2023): 42–47.
9. H. R. Xia, Z. Wang, Y. Wang, Z. Gao, and M. Xiao. "Fully flat bands in a photonic dipolar Kagome lattice." *Phys Rev Lett* 135, no. 17 (2025): 176902.
10. P. Ren, J. Zheng, Z. Huang, Y. Liu, L. Zhang, H. Zhang, J. Ma, Z. Chen, J. F. Li, J. Yi, and Z. Yang. "Far-field excitation of a photonic flat band via a tailored anapole mode." *Phys Rev Lett* 135, no. 8 (2025): 083803.
11. J. Si, Z. Zhang, H. Wu, H. Wang, J. Wang, J. Liu, M. Qin, and H. Li. "Flat-band polaritons in van der Waals metasurfaces driven by collective coupling of brillouin-zone-folded Q-BIC, anapoles, and excitons." *Photonics Res* 14, no. 8 (2026): 3482–3496.
12. H. Tang, F. Du, S. Carr, C. DeVault, O. Mello, and E. Mazur. "Modeling the optical properties of twisted bilayer photonic crystals." *Light Sci Appl* 10, no. 1 (2021): 157.
13. K. Dong, T. Zhang, J. Li, Q. Wang, F. Yang, Y. Rho, D. Wang, C. P. Grigoropoulos, J. Wu, and J. Yao. "Flat bands in magic-angle bilayer photonic crystals at small twists." *Phys Rev Lett* 126, no. 22 (2021): 223601.
14. D. Choi, S. J. Lee, S. W. Cho, C. B. Bark, S. Pak, D. J. Shin, M. J. Park, and S. H. Gong. "Strong interlayer coupling and chiral flat-band cascades in twisted bilayer gratings." *Sci Adv* 12, no. 35 (2026): eaed8846.
15. S. Noda, M. Fujita, and T. Asano. "Spontaneous-emission control by photonic crystals and nanocavities." *Nat Photonics* 1, no. 8 (2007): 449–458.
16. E. A. Muljarov and W. Langbein. "Exact mode volume and Purcell factor of open optical systems." *Phys Rev B* 94, no. 23 (2016): 235438.
17. T. X. Hoang, A. Nussupbekov, J. Ji, D. Leykam, J. Gómez Rivas, and Y. Kivshar. "Mie-tronics supermodes and symmetry breaking in nonlocal metasurfaces." *Phys Rev Res* 8, no. 3 (2026): 033076.
18. W. Chen, Y. Chen, and W. Liu. "Multipolar conversion induced subwavelength high-q Kerker supermodes with unidirectional radiations." *Laser Photonics Rev* 13, no. 9 (2019): 1900067.
19. O. Khatib, S. Ren, J. Malof, and W. J. Padilla. "Learning the Physics of All-Dielectric Metamaterials with Deep Lorentz Neural Networks." *Adv Opt Mater* 10, no. 13 (2022): 2200097.
20. J. Fan, Y. Zhou, Z. Xue, G. Xu, J. Chen, H. Xing, and L. Cong. "High-efficiency active membrane metasurfaces." *Sci Adv* 11, no. 50 (2025): eadw4752.
21. A. C. Overvig, S. C. Malek, and N. Yu. "Multifunctional nonlocal metasurfaces." *Phys Rev Lett* 125, no. 1 (2020): 017402.
22. Y. Zhou, S. Guo, A. C. Overvig, and A. Alù. "Multiresonant nonlocal metasurfaces." *Nano Lett* 23, no. 14 (2023): 6768–6775.
23. B. Auguié and W.L. Barnes. "Collective Resonances in Gold Nanoparticle Arrays." *Phys. Rev. Lett.* 101, no. 14 (2008): 143902.
24. G. Vecchi, V. Giannini, and J. Gómez Rivas. "Shaping the Fluorescent Emission by Lattice Resonances in Plasmonic Crystals of Nanoantennas." *Phys. Rev. Lett.* 102, no. 14 (2009): 146807.
25. M. Manjappa, P. Pitchappa, N. Singh, N. Wang, N. I. Zheludev, C. Lee, and R. Singh. "Reconfigurable MEMS Fano metasurfaces with multiple-input–output states for logic operations at terahertz frequencies." *Nat Commun* 9, no. 1 (2018): 4056.
26. Y. Liang, D. P. Tsai, and Y. Kivshar. "From local to nonlocal high-q plasmonic metasurfaces." *Phys Rev Lett* 133, no. 5 (2024): 053801.
27. T. X. Hoang, D. Leykam, H.-S. Chu, C. E. Png, F. J. García-Vidal, and Y. S. Kivshar. "Collective nature of high-Q resonances in finite-size photonic metastructures." *Phys Rev Res* 7, no. 1 (2025): 013316.
28. W. Lv, H. Qin, X. Shi, F. Shi, X. Mu, Z. Zhou, J. Qin, B. Li, C. W. Qiu, and Q. Song. "Local-nonlocal assisted multifunctional photonic crystals." *Light Sci Appl* 15, no. 1 (2026): 243.
29. L. M. Máñez-Espina and A. Díaz-Rubio. "Control of the local and nonlocal electromagnetic response in all-dielectric reconfigurable metasurfaces." *Phys Rev Appl* 22, no. 5 (2024): 054022.
30. B. Zhang, S. Li, A. Dong, M. Qin, L. Huang, L. Li, and H. Li. "High-Q chiral flat-band guided resonances in nonlocal metagratings." *Nano Lett* 26, no. 17 (2026): 5977–5984.
31. V. V. Kumar and E. Semouchkina. "Enhancing resonance wave–matter interaction in finite-size dielectric metasurfaces." *Journal of Optics* 28, no. 7 (2026): 075403.
32. A. Silva, F. Monticone, G. Castaldi, V. Galdi, A. Alù, and N. Engheta. "Performing mathematical operations with metamaterials." *Science* 343, no. 6167 (2014): 160–163.
33. F. Zangeneh-Nejad, D. L. Sounas, A. Alù, and R. Fleury. "Analogue computing with metamaterials." *Nat Rev Mater* 6, no. 3 (2021): 207–225.
34. H. Yuan, X. Zhao, J. Zhang, J. Liang, W. Peng, H. Xing, G. Xu, C. Zhang, J. Lou, and L. Cong. "3-Bit Parallel Logic Operations for Ultrafast Computing Based on Light-Driven Terahertz Metadevice." *Adv Funct Mater* 36, no. 15 (2026): e17748.

35. G. Lozano, G. Grzela, M.A. Verschuuren, M. Ramezani, and J. Gómez Rivas. "Tailor-made directional emission in nanoimprinted plasmonic-based light-emitting devices." *Nanoscale* 6, no. 15 (2014): 9223.
36. J. H. Song, J. van de Groep, S. J. Kim, and M. L. Brongersma. "Non-local metasurfaces for spectrally decoupled wavefront manipulation and eye tracking." *Nat Nanotechnol* 16, no. 11 (2021): 1224–1230.
37. Y. Wang, T. Wang, M. Xia, K. Ding, Y. Liao, Y. Chen, and X. Zhang. "Steering chiral exciton emission with all van der Waals nonlocal phase gradient metasurfaces." *Adv Funct Mater* 36, no. 89 (2026): e78719.
38. R. Premanand, Y. J. Tan, T. C. Tan, P. Pitchappa, H. K. Cheemalamarri, A. S. Madhukumar, and R. Singh. "Broadband Silicon-On-Glass Reflective Dielectric Metasurface for THz Beam Steering in 6G Wireless Networks." *Adv Funct Mater* 36, no. 75 (2026): e77642.
39. A. I. Kuznetsov, A. E. Miroshnichenko, M. L. Brongersma, Y. S. Kivshar, and B. Luk'yanchuk. "Optically resonant dielectric nanostructures." *Science* 354, no. 6314 (2016): aag2472.
40. N. Rozman, K. Fan, and W. J. Padilla. "Symmetry-broken high-Q terahertz quasi-bound states in the continuum." *ACS Photonics* 11, no. 5 (2024): 1893–1900.
41. S. Long, W. Zhang, Z. Xue, G. Xu, P. P. Shum, D. Luo, and L. Cong. "Active Terahertz Nonlocal Metasurfaces With Liquid Crystal Elastomers." *Laser Photonics Rev* 19, no. 13 (2025): 2402167.
42. H. Qiu, X. Tu, M. Qin, F. Wu, T. Liu, and S. Xiao. "Electro-optically tunable second-harmonic generation in lithium niobate metasurfaces boosted by Brillouin zone folding induced bound states in the continuum." *Phys Rev B* 112, no. 8 (2025): 085418.
43. B. Munkhbat, B. Küçüköz, D. G. Baranov, T. J. Antosiewicz, and T. O. Shegai. "Nanostructured Transition Metal Dichalcogenide Multilayers for Advanced Nanophotonics." *Laser Photonics Rev* 17, no. 1 (2023): 2200057.
44. V. Karimi, D. Bosomtwi, L. Liu, and V. E. Babicheva. "Enhancing MXene Multipolar Resonances Through Metasurface Lattice Engineering." *Nano Select* 6, no. 9 (2025): e70002.
45. X. Zhang. "From anapole to polariton: A fundamental transition in the mechanism underlying robust flatband generation in WS2 metasurfaces." *Physical Review B* 113, no. 3 (2026): 035401.
46. G. W. Castellanos, P. Bai, and J. Gómez Rivas. "Lattice resonances in dielectric metasurfaces." *J Appl Phys* 125, no. 21 (2019): 213105.
47. A. V. Prokhorov, P. D. Terekhov, M. Y. Gubin, A. V. Shesterikov, X. Ni, V. R. Tuz, and A. B. Evlyukhin. "Resonant light trapping via lattice-induced multipole coupling in symmetrical metasurfaces." *ACS Photonics* 9, no. 12 (2022): 3869–3875.
48. I. Allayarov, V. Aita, D. J. Roth, B. van Casteren, A. Y. Bykov, A. B. Evlyukhin, A. V. Zayats, and A. Calà Lesina. "Strong coupling of collective optical resonances in dielectric metasurfaces." *Light Sci Appl* 14, no. 1 (2025): 387.
49. L. Feng and X. Zhang. "Beyond Fourier harmonics: anapole-engineered flat-band quasi–bound states in the continuum in dielectric metasurfaces." *Phys Rev B* 112, no. 11 (2025): 115442.
50. Guizhen Xu, Hongyang Xing, Zhanqiang Xue, Dan Lu, Jinying Fan, Junxing Fan, Perry Ping Shum, and Longqing Cong. "Recent advances and perspective of photonic bound states in the continuum." *Ultrafast Science* 3, no. 5 (2023): 0033.
51. L. Hidalgo-Arteaga, J. L. Pura, B. García-Cámara, A. Barreda, and J.A. Sánchez-Gil. "Narrow-band asymmetric THz absorbers and polarizers based on impedance-matched, cloaked quasi-BICs in doped Ge microdisk metasurfaces." *Iscience* 29, no. 6 (2026): 116169.
52. W. Wang, Z. Shen, and R. Singh. "Chip-Scale Terahertz Sensing Driven by Supercoupling Induced Transparency in Topological Cavities." *Laser Photonics Rev* xx, no. x (2026): e72140.
53. Z. Xue, H. Xing, G. Xu, J. Fan, Z. Liao, Y. Zhou, and L. Cong. "High-Density Pixelated Metasurfaces for Terahertz Hyperspectral Imaging." *Adv Mater* xx, no. x (2026): e75661.
54. N. Navaratna, T. C. Tan, P. Szriftgiser, G. Ducournau, and R. Singh. "Ultralow-Dispersion Silicon THz Photonic Crystal Interconnects for Multichannel Datalinks." *Adv Mater* 38, no. 48 (2026): e74150.
55. R. Heilmann, A. I. Väkeväinen, J.-P. Martikainen, and P. Törmä. "Strong coupling between organic dye molecules and lattice modes of a dielectric nanoparticle array." *Nanophotonics* 9, no. 2 (2020): 267–276.
56. V. E. Babicheva and A. B. Evlyukhin. "Mie-resonant metaphotonics." *Adv Opt Photonics* 16, no. 3 (2024): 539–658.
57. S. R. K. Rodriguez, M. C. Schaafsma, A. Berrier, and J. G. Rivas. "Collective resonances in plasmonic crystals: size matters." *Physica B Condens Matter* 407, no. 20 (2012): 4081–4085.
58. T. X. Hoang, D. Leykam, and Y. Kivshar. "Photonic flatband resonances in multiple light scattering." *Phys Rev Lett* 132, no. 4 (2024): 043803.
59. J. Yang, Y. Li, Y. Yang, X. Xie, Z. Zhang, J. Yuan, H. Cai, Da W. Wang, and F. Gao. "Realization of all-band-flat photonic lattices." *Nat Commun* 15, no. 1 (2024): 1484.
60. Z. Li, S. Yu, and X. Ao. "Dual flat bands enable synergistic excitation–Emission enhancement in rare-earth ions." *ACS Photonics* 13, no. 19 (2026): 5780–5786.
61. Q. Wang, S. Stobbe, and P. Lodahl. "Mapping the local density of optical states of a photonic crystal with single quantum dots." *Phys Rev Lett* 107, no. 16 (2011): 167404.
62. C. P. Mavidis, A. C. Tasolamprou, S. B. Hasan, T. Koschny, E. N. Economou, M. Kafesaki, C. M. Soukoulis, and W. L. Vos. "Local density of optical states in the three-dimensional band gap of a finite photonic crystal." *Phys Rev B* 101, no. 23 (2020): 235309.
63. W. L. Barnes, S. A. R. Horsley, and W. L. Vos. "Classical antennas, quantum emitters, and densities of optical states." *J Opt* 22, no. 7 (2020): 073501.
64. S. E. T. ter Huurne, D. B. L. Peeters, J. A. Sánchez-Gil, and J. Gómez Rivas. "Direct measurement of the local density of optical states in the time domain." *ACS Photonics* 10, no. 8 (2023): 2980.
65. J. Ji, J. A. Sánchez-Gil, D. Peeters, W. Holman, T. X. Hoang, D. van Mechelen, and J. Gómez Rivas. "Near-field probing of the local density of optical states enhanced by bound states in the continuum in nonlocal metasurfaces." *Nat Commun* 16, no. 1 (2025): 11597.
66. L. Novotny and B. Hecht. 2012. *Principles of Nano-Optics*, Cambridge: Cambridge University Press.
67. R. Sapienza, T. Coenen, J. Renger, M. Kuttge, N. F. Van Hulst, and A. Polman. "Deep-subwavelength imaging of the modal dispersion of light." *Nat Mater* 11, no. 9 (2012): 781–787.
68. I. S. Nikolaev, W. L. Vos, and A. F. Koenderink. "Accurate calculation of the local density of optical states in inverse-opal photonic crystals." *J Opt Soc Am B* 26, no. 5 (2009): 987–997.
69. P. Qiu, G. Wang, J. Lu, and H. Wang. "Local density of states in photonic crystal cavity." *Front Optoelectron* 5, no. 3 (2012): 341–344.
70. A. F. Koenderink. "On the use of Purcell factors for plasmon antennas." *Opt Lett* 35, no. 24 (2010): 4208–4210.
71. P. Lodahl, S. Mahmoodian, and S. Stobbe. "Interfacing single photons and single quantum dots with photonic nanostructures." *Rev Mod Phys* 87, no. 2 (2015): 347.
72. V. E. Babicheva and A. B. Evlyukhin. "Metasurfaces with electric quadrupole and magnetic dipole resonant coupling." *ACS Photonics* 5, no. 5 (2018): 2022–2033.
73. Z. Sadrieva, K. Frizyuk, M. Petrov, Y. Kivshar, and A. Bogdanov. "Multipolar origin of bound states in the continuum." *Phys Rev B* 100, no. 11 (2019): 115303.
74. S. Murai, G. W. Castellanos, T. V. Raziman, A. G. Curto, and J. Gómez Rivas. "Enhanced light emission by magnetic and electric resonances in dielectric metasurfaces." *Adv Opt Mater* 8, no.

16 (2020): 1902024.
75. Y. Yang, I. I. Kravchenko, D. P. Briggs, and J. Valentine. "All-dielectric metasurface analogue of electromagnetically induced transparency." *Nat Commun* 5, no. 1 (2014): 5753.
76. A. Barreda, A. García-Martín, and J. A. Sánchez-Gil. "Bound states in the continuum in all-dielectric metasurfaces." *APL Photonics* 10, no. 10 (2025): 100902.
77. N. Ustimenko, C. Rockstuhl, and A. B. Evlyukhin. "Resonances in finite-size all-dielectric metasurfaces for light trapping and propagation control." *Phys Rev B* 109, no. 11 (2024): 115436.
78. T. Hoekstra, S. A. Mann, and J. van de Groep. "Finite-size effects in nonlocal metasurfaces." *ACS Photonics* 13, no. 18 (2026): 5337–5346.
79. W. J. Holman, J. Ji, and J. Gómez Rivas. "Extended nonradiative dipole-dipole energy transfer and spatiotemporal coherence enabled by bound states in the continuum." *Sci Adv* 12, no. 18 (2026): 1–9.
80. W. J. Holman, J. Ji, and J. Gómez Rivas. "THz near-field spatial mapping of dipole–dipole interactions in free space and near interfaces." *J Appl Phys* 139, no. 20 (2026): 203104.
81. A. Vaskin, R. Kolkowski, A. F. Koenderink, and I. Staude. "Light-emitting metasurfaces." *Nanophotonics* 8, no. 7 (2019): 1151–1198.
82. L. Heki, Y. Mohtashami, R. A. Decrescent, A. Alhassan, S. Nakamura, S. P. Denbaars, and J. A. Schuller. "Designing highly directional luminescent phased-array metasurfaces with reciprocity-based simulations." *ACS Omega* 7, no. 26 (2022): 22477–22483.
83. K. J. Vahala. "Optical microcavities." *Nature* 424, no. 6950 (2003): 839–846.
84. P. T. Kristensen and S. Hughes. "Modes and mode volumes of leaky optical cavities and plasmonic nanoresonators." *ACS Photonics* 1, no. 1 (2014): 2–10.

## Supporting Information

Additional supporting information can be found online in the Supporting Information section.

Supporting Information

# Observation of collective flatbands in hybrid Mie-grating THz metasurfaces

Jie Ji[1*], Thanh Xuan Hoang[2†], Ayan Nussupbekov[3], Siyan Zhou[4], Kehan Wang[1], Yuri Kivshar[5*], Jaime Gómez Rivas[1*]

*E-mail:j.ji@tue.nl,yuri.kivshar@anu.edu.au, j.gomez.rivas@tue.nl.

## Contents

## S1 Fabrication process of the free-standing silicon membranes

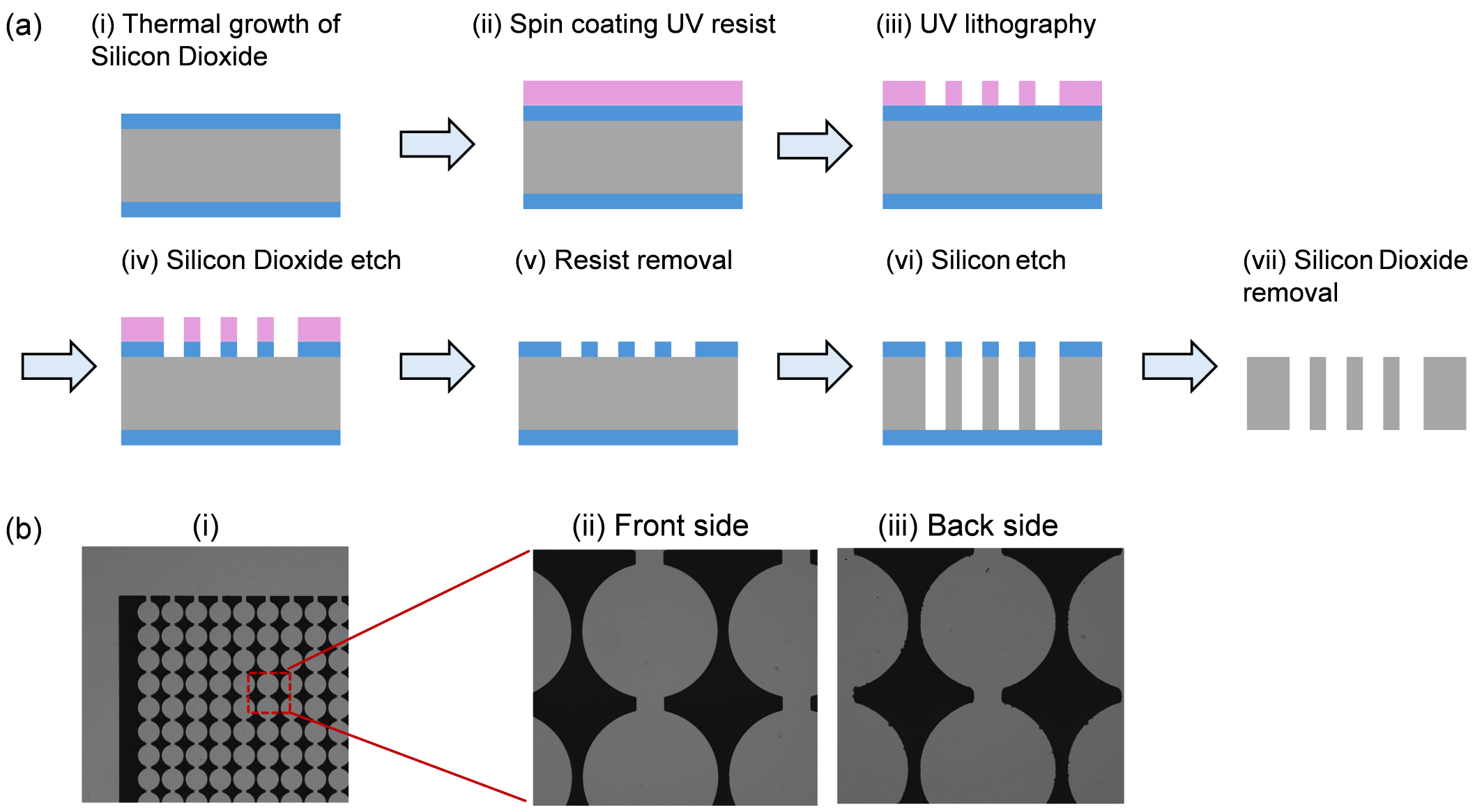

Figure S1: **Fabrication process of the free-standing silicon membrane and optical image of the fabricated samples.** (a) Schematic illustration of the fabrication process for the free-standing silicon membrane. (i) Thermal growth of a $SiO_2$ layer on the substrate. (ii) Spin-coating of UV photoresist. (iii) Pattern definition by UV lithography. (iv) Transfer of the pattern into silicon dioxide via etching. (v) Removal of the residual photoresist. (vi) Transfer of the pattern into silicon via etching. (vii) Removal of the $SiO_2$ layer, leaving the final silicon grating array. (b) Optical images of the fabricated membrane, showing a periodic pattern of circular features connected with bridges and forming a hybrid Mie–grating metasurface; the red dashed box indicates the region selected for magnified inspection. (ii) Enlarged view of the front side, revealing the etched structures with well-defined geometries. (iii) Corresponding back-side view, confirming pattern transfer and structural symmetry through the wafer.

Figure S1 shows the fabrication process flow of the free-standing silicon membranes. Silicon wafers were first subjected to a standard RCA cleaning procedure to eliminate organic, ionic, and metallic contaminants. This cleaning step ensured a pristine surface for subsequent processing. Following the cleaning, thermal oxidation was performed in an anneal-oxide furnace at 1100 $°C$ for 3 hours under a water vapor-rich ambient to promote wet oxidation. This process resulted in the formation of a high-quality silicon dioxide ($SiO_2$) layer with a uniform thickness of approximately 1050 nm. Then, patterns of the metasurface were fabricated using a standard photolithography process. AZ 5214E positive photoresist was spin-coated onto the oxidized wafers to achieve a uniform thickness of 2 $\mu m$. Pattern exposure was carried out using a 375 nm laser direct writing system (MLA 150, Heidelberg). Subsequently, the samples were developed in AZ 726 MIF developer for 1 minute.

The etching of the $SiO_2$ layer was performed using a Deep Reactive Ion Etching (DRIE) system (SPTS Pegasus). The process parameters were optimized for high anisotropy, with a chamber pressure maintained at 8.8 mTorr. The gas mixture included $C_4F_8$, $H_2$, and He, with a $He/C_4F_8$ flow ratio of 17.5 and a $C_4F_8/H_2$ ratio of 1, ensuring selective and vertical etching profiles. After oxide etching, residual photoresist was completely removed via oxygen plasma ashing in a Plasma Asher system. The etching process was conducted with an oxygen flow of 500 sccm for 1 hour. Silicon etching was subsequently carried out using the Bosch DRIE process on the same SPTS Pegasus platform. This time-multiplexed process alternates between passivation and etching cycles to achieve high aspect ratio structures with steep sidewalls. During the passivation step, a protective fluorometer layer was deposited using $C_4F_8$ gas. The etching step then selectively removed this layer from the trench bottoms while etching the silicon using $SF_6$ gas, enabling deep and anisotropic etching.

Lastly, the remaining $SiO_2$ layers were removed from both wafer surfaces using a Buffered Hydrofluoric Acid (BHF) bath. The wafers were immersed in BHF for half an hour to ensure complete oxide removal. Following the etch, wafers were rinsed thoroughly with deionized water using a bubbler system and dried with nitrogen blow-drying or a spin-rinse dryer, completing the fabrication process and preparing the samples for subsequent analysis.

Figure S1(b) shows optical microscope images of the fabricated patterns. As shown in Figure S1(b)(i), the free-standing array exhibits a high degree of uniformity over a large area and remains connected to the bulk substrate via the supporting grating structure. A magnified view of a selected region [Figure S1(b)(ii)] reveals well-defined circular features with smooth edges and consistent spacing. The corresponding back-side image [Figure S1(b)(iii)] confirms successful through-wafer etching, with no observable distortion or collapse of the high-aspect-ratio features, and demonstrates good agreement with the front-side morphology. These results verify that the employed fabrication process enables precise realization of the designed metasurface geometry.

## S2 Multipolar analysis of hybrid modes in the grating-free metasurface

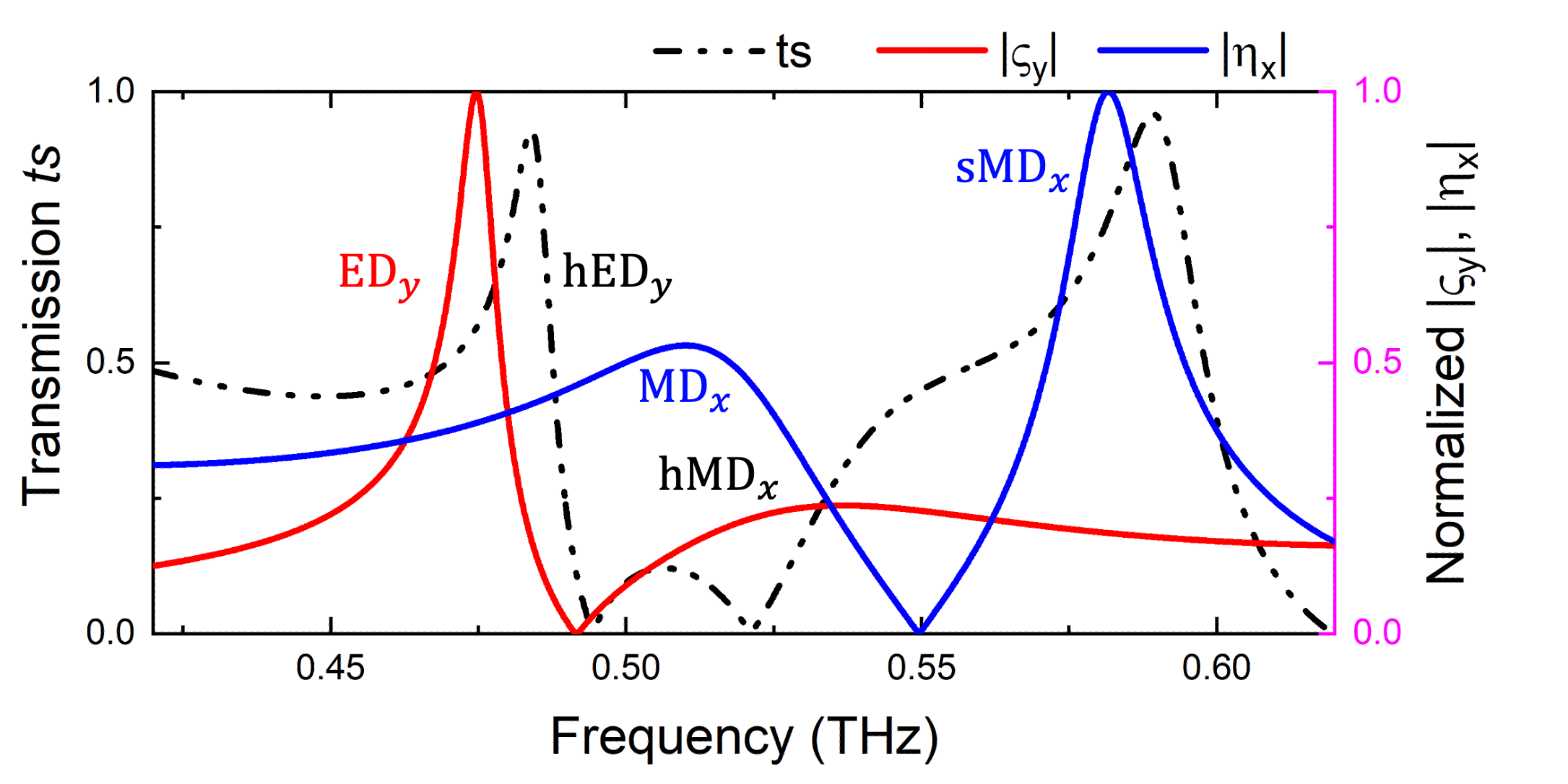

Figure S2: Transmission spectrum under TE-polarized normal incidence ($\mathbf{k}_z, \mathbf{E}_y, \mathbf{H}_x$) for the cylindrical Mie-resonator array, showing the two hybrid modes $\mathrm{hED_y}$ and $\mathrm{hMD_x}$ discussed in the main text. The right $y$ axis shows the normalized electric- and magnetic-dipole expansion coefficients, $|\zeta_y|$ and $|\eta_x|$, respectively, representing the internal fields. Three principal dipolar resonances contribute to the spectral response: a $y$-oriented electric dipole ($\mathrm{ED_y}$), an $x$-oriented magnetic dipole ($\mathrm{MD_x}$), and a higher-frequency single-dominant $x$-oriented magnetic-dipole mode ($\mathrm{sMD_x}$). Within the frequency range $0.45$–$0.55$ THz, the $\mathrm{ED_y}$ and $\mathrm{MD_x}$ resonances are simultaneously excited and hybridize to form the two hybrid modes $\mathrm{hED_y}$ and $\mathrm{hMD_x}$.

To investigate the role of the dielectric interconnections in multipolar engineering, we first analyze the multipolar composition of the eigenmodes of the metasurface without grating under TE-polarized normal incidence. Figure S2 shows an extended transmission spectrum, with the two hybrid modes $\mathrm{hED_y}$ and $\mathrm{hMD_x}$ discussed in the main text labeled for ease of reference. In general, the spectral response is governed by the interaction of the incident plane wave with a small number of low-order multipolar resonances, most prominently with dipolar contributions. Resonant scattering commonly produces asymmetric Fano-like spectral features through interference between resonant and non-resonant pathways. Such features are evident in Figure S2, where three principal dipolar resonances contributing to the transmission response are identified: $\mathrm{ED_y}$, $\mathrm{MD_x}$, and $\mathrm{sMD_x}$.

Before providing a detailed multipolar analysis of the three modes, it is instructive to con-sider the multipole expansion of the electromagnetic field induced by an incident field. For the lowest-order electric and magnetic dipoles, the Cartesian expansion coefficients can be related directly to the corresponding internal-field components at the expansion origin as

$$\zeta_x = \sqrt{\frac{3}{8\pi}}\,\lambda E_x, \qquad \zeta_y = \sqrt{\frac{3}{8\pi}}\,\lambda E_y, \qquad \zeta_z = \sqrt{\frac{3}{8\pi}}\,\lambda E_z, \tag{1}$$

and

$$\eta_x = \sqrt{\frac{3}{8\pi}}\,\lambda H_x, \qquad \eta_y = \sqrt{\frac{3}{8\pi}}\,\lambda H_y, \qquad \eta_z = \sqrt{\frac{3}{8\pi}}\,\lambda H_z, \tag{2}$$

where $\zeta_\alpha$ and $\eta_\alpha$ ($\alpha = x, y, z$) are the Cartesian electric- and magnetic-dipole expansion coefficients, respectively, and $\lambda$ is the wavelength in the corresponding material. These linear relations provide a direct connection between the dipolar expansion coefficients and the Cartesian components of the internal electromagnetic field, making them particularly convenient for identifying and quantifying the electric- and magnetic-dipole contributions to the resonant modes. Accordingly, the right $y$ axis in Figure S2 shows the normalized spectra of the two relevant Cartesian dipolar coefficients, $|\zeta_y|$ and $|\eta_x|$, extracted from the simulated near-field distributions using Eqs. 1 and 2.

Even though our main interest lies in the diffraction bands associated with the hybrid modes $\mathrm{hED_y}$ and $\mathrm{hMD_x}$, which arise from lattice-mediated hybridization between the two principal resonances $\mathrm{ED_y}$ and $\mathrm{MD_x}$ and manifest as Fano-like features through their interference with the incident-field background, it is instructive to first examine the mode profile of $\mathrm{sMD_x}$. In contrast to these hybrid modes, $\mathrm{sMD_x}$ is dominated by a single magnetic-dipole resonance, providing a useful reference for distinguishing a single-dominant mode from hybrid multipolar states.

Figure S3 presents all six Cartesian field components associated with the $\mathrm{sMD_x}$ response. The field patterns of Im($E_x$), Im($E_y$), and Re($H_z$) are weak and predominantly represent direct off-resonant excitation of a grating-like background. In particular, Im($E_y$) follows the polarization of the incident wave and exhibits a comparatively featureless diffraction-like distribution, while the corresponding Im($E_x$) and Re($H_z$) components remain similarly weak and poorly defined. In contrast, Re($H_x$), Re($H_y$), and Im($E_z$) exhibit well-defined spatial patterns characteristic of the resonant $\mathrm{sMD_x}$ mode, arising from resonance-induced displacement-current circulation. The transmission response near the $\mathrm{sMD_x}$ resonance is therefore largely governed

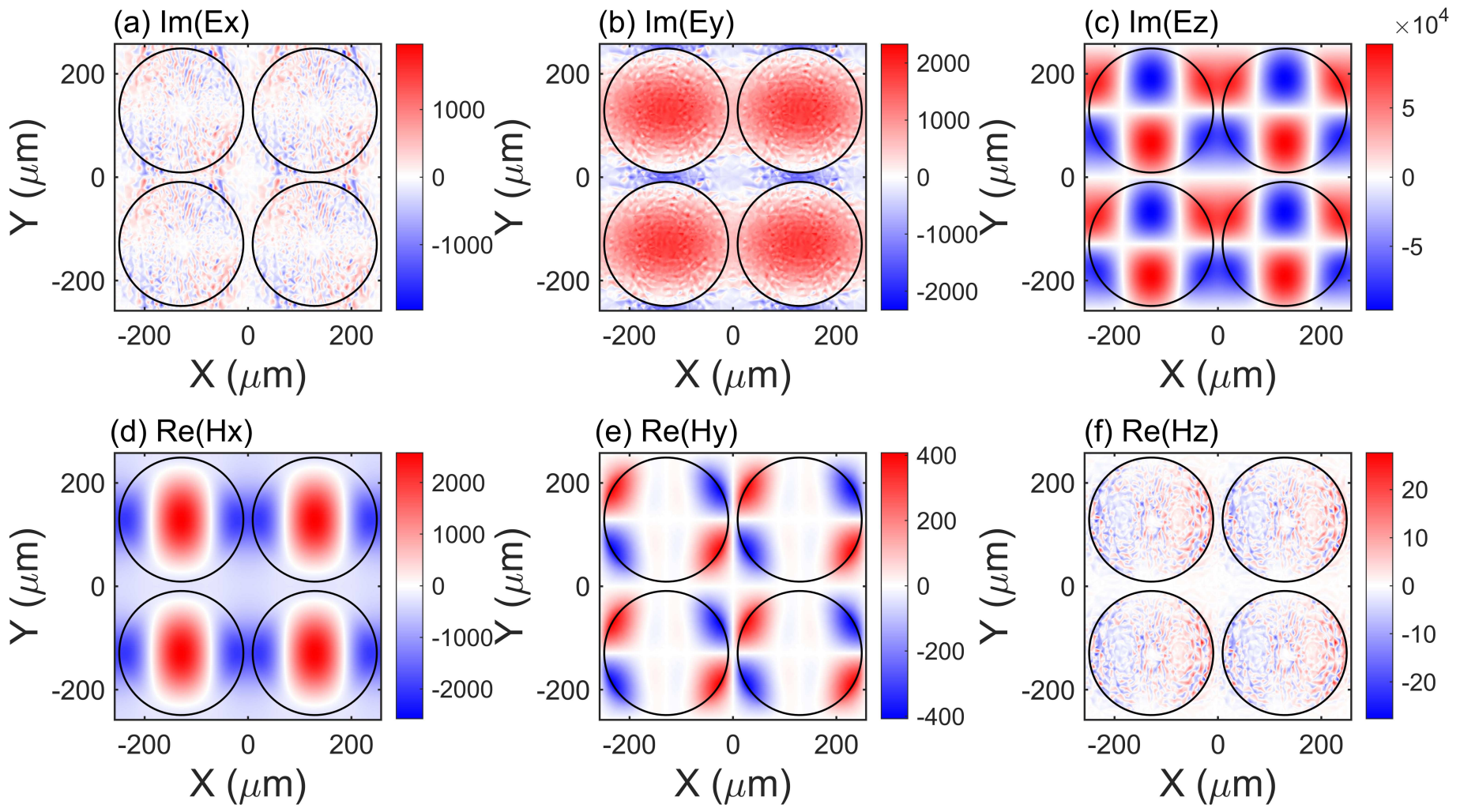

Figure S3: **Mode profile of the single-dominant $x$-oriented magnetic-dipole mode $\mathrm{sMD_x}$ under TE-polarized normal incidence ($\mathbf{k}_z, \mathbf{E}_y, \mathbf{H}_x$).** (a–c) Imaginary parts of the electric-field components, Im($E_x$), Im($E_y$), and Im($E_z$), respectively. (d–f) Real parts of the magnetic-field components, Re($H_x$), Re($H_y$), and Re($H_z$), respectively. The well-defined field patterns of Re($H_x$), Re($H_y$), and Im($E_z$) exhibit the characteristic symmetry of the resonant $\mathrm{sMD_x}$ mode. In contrast, Im($E_x$), Im($E_y$), and Re($H_z$) arise predominantly from direct off-resonant excitation of a grating-like background and are weaker than the resonant $\mathrm{sMD_x}$ contribution by orders of magnitude.

by a single dominant magnetic-dipole mode.

Such single-mode dominance provides fewer modal degrees of freedom for controlling the spectral response. Hybridization between multiple resonant multipoles introduces additional interference and coupling channels, thereby expanding the accessible design space. This distinction is illustrated in the $0.45$–$0.55$ THz range, where the $\mathrm{ED_y}$ and $\mathrm{MD_x}$ resonances hybridize to form the two modes $\mathrm{hED_y}$ and $\mathrm{hMD_x}$. Their relative spectral positions and interference can then be engineered through geometric modification of the metasurface, providing a richer route for controlling the resulting nonlocal response.

Figure S4 shows the mode profiles of the two hybrid modes at the resonance frequencies associated with the $\mathrm{ED_y}$ and $\mathrm{MD_x}$ features identified in Figure S2. Figure S4(a) shows that, although the $\mathrm{ED_y}$ contribution is more than one order of magnitude stronger than the $\mathrm{MD_x}$ contribution, the weaker magnetic-dipole component remains well defined. This coexistence of the two resonant contributions highlights the hybrid nature of the $\mathrm{hED_y}$ mode. We note that the $\mathrm{ED_y}$ and $\mathrm{MD_x}$ modes are orthogonal for an isolated resonator; however, lattice-mediated coupling enables their interaction in the periodic metasurface, giving rise to the hybrid modes $\mathrm{hED_y}$ and $\mathrm{hMD_x}$.

Figure S4(b) reveals a different hybridization regime. Here, the $\mathrm{MD_x}$ contribution is dominant and exhibits a well-defined resonant field pattern, whereas the $\mathrm{ED_y}$ contribution shows a pronounced signature of destructive interference, most clearly visible in the spatial distribution of Re($H_z$). Two distinct features can be identified: one localized predominantly within the unit cell, similar to that observed in Figure S4(a,vi), and another extending across the unit-cell boundaries. The former may be associated with a Mie-type contribution, whereas the latter is characteristic of a grating-type contribution because of its spatially extended character across neighboring unit cells. Their coexistence and interference strongly suppress the net $\mathrm{ED_y}$ contribution in the $\mathrm{hMD_x}$ hybrid mode. The destructive interference is manifested by the opposite signs of Re($H_z$) across the unit-cell boundary, corresponding to approximately out-of-phase oscillations of the Mie-type and grating-type contributions.

A further important distinction between the two constituent resonances lies in their displacement-current circulation. The displacement current associated with $\mathrm{MD_x}$ circulates predominantly in the $yz$ plane, whereas that associated with $\mathrm{ED_y}$ circulates predominantly in the $xz$ plane. In the following section, we show that this distinct spatial character enables a simple yet powerful design strategy based on interconnections between the adjacent unit cells.

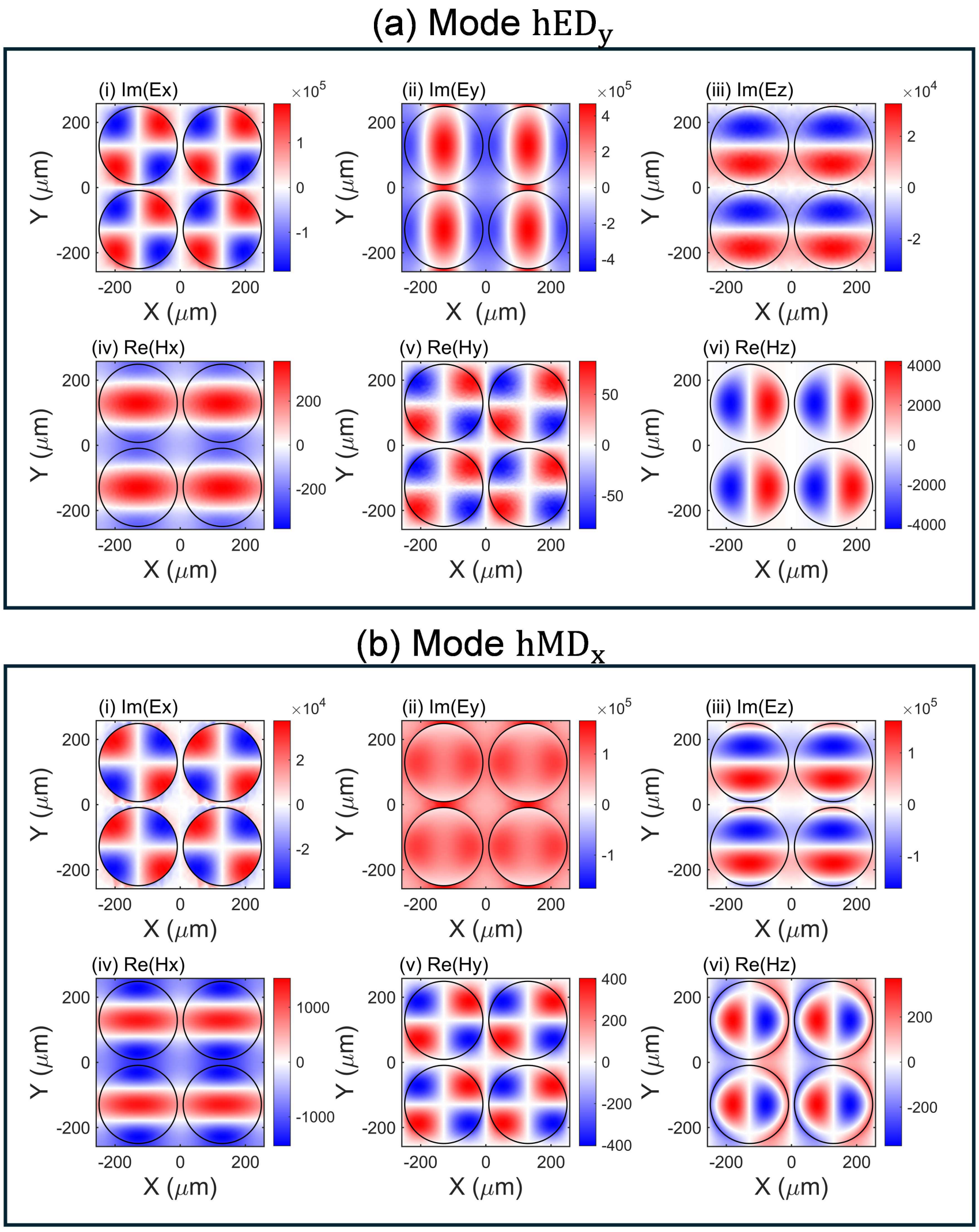

Figure S4: **Multipolar analysis of the hybrid modes** $\mathrm{hED_y}$ **and** $\mathrm{hMD_x}$ **under TE-polarized normal incidence ($\mathbf{k}_z, \mathbf{E}_y, \mathbf{H}_x$).** (a) Mode profile of $\mathrm{hED_y}$, showing signatures of both the resonant $\mathrm{ED_y}$ and $\mathrm{MD_x}$ modes, with stronger $\mathrm{ED_y}$ character. The $\mathrm{ED_y}$ contribution is represented by Im($E_x$), Im($E_y$), and Re($H_z$), whereas the $\mathrm{MD_x}$ contribution is represented by Re($H_x$), Re($H_y$), and Im($E_z$). (b) Mode profile of $\mathrm{hMD_x}$, showing a pronounced $\mathrm{MD_x}$ character, while the $\mathrm{ED_y}$ contribution is strongly suppressed.

# S3 Grating-width-controlled multipolar engineering of the hybrid modes

Having established the multipolar composition of the hybrid modes in the metasurface formed by cylindrical dielectric resonators, we now examine how the grating modifies the constituent resonances and their hybridization. We first consider a $50$ $\mu$m-wide grating and, subsequently, trace the evolution of the resonances as the grating width is varied.

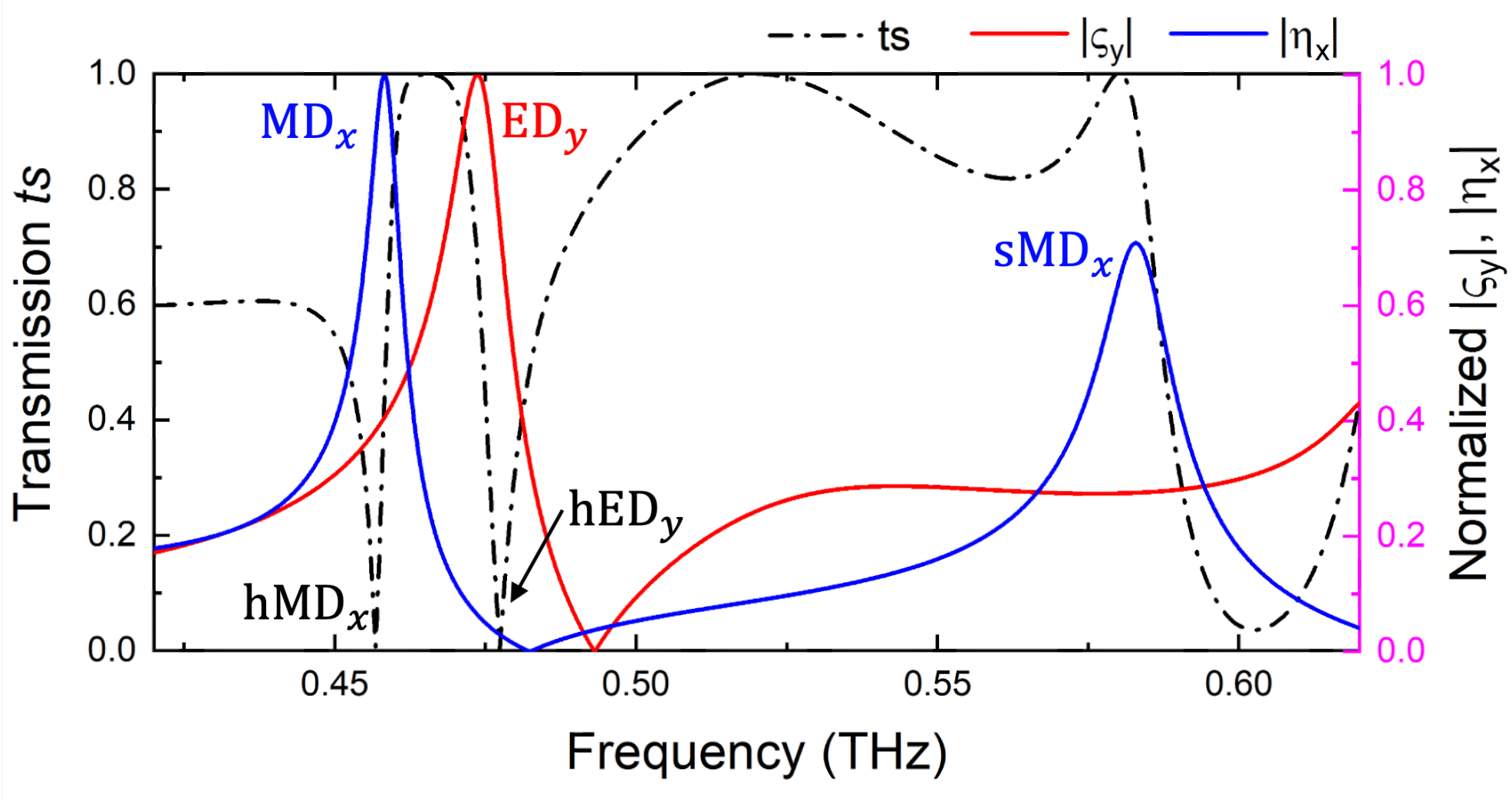

Figure S5: Transmission spectrum under TE-polarized normal incidence ($\mathbf{k}_z, \mathbf{E}_y, \mathbf{H}_x$) for the metasurface with a $50$ $\mu$m-wide $y$-oriented grating, showing the reversed spectral ordering of the hybrid modes $\mathrm{hED_y}$ and $\mathrm{hMD_x}$ compared with the grating-free case in Figure S2.

Figure S5 illustrates the effect of introducing a $50$ $\mu$m-wide grating oriented along the $y$ direction. Compared to the grating-free metasurface shown in Figure S2, the grating shifts the $\mathrm{MD_x}$ resonance from the high-frequency side to the low-frequency side of the $\mathrm{ED_y}$ resonance, thereby reversing their spectral ordering. At the same time, the spectral profile of the hybrid $\mathrm{hMD_x}$ mode becomes substantially narrower. This pronounced spectral shift can be understood from the spatial overlap between the interconnection region and the displacement-current distributions of the constituent resonances. Because the grating is oriented along the $y$ direction, the displacement current associated with $\mathrm{MD_x}$, which circulates predominantly in the $yz$ plane, has strong spatial overlap with the interconnection region of the grating. In contrast, the $\mathrm{ED_y}$ current circulates predominantly in the $xz$ plane and therefore overlaps only weakly with the dielectric interconnection region. As a result, the $\mathrm{MD_x}$ resonance is shifted substantially by the grating, whereas the $\mathrm{ED_y}$ resonance remains comparatively unchanged.

It is also noteworthy that the $\mathrm{sMD_x}$ mode shown in Figure S3 likewise exhibits displacement-current circulation predominantly in the $yz$ plane, yet its resonance frequency changes only weakly upon introducing the grating [Figure S5]. This difference arises from its distinct spatial field distribution: the fields associated with $\mathrm{sMD_x}$ nearly vanish at the interconnection region of the grating, resulting in much weaker modal overlap with the interconnection region than for the $\mathrm{MD_x}$ resonance.

Figure S6 further elucidates the effect of the grating through the modified multipolar composition of the two hybrid modes. For the $\mathrm{hMD_x}$ mode in Figure S6(a), compared with the grating-free case in Figure S4(b), the $\mathrm{ED_y}$ contribution represented by Im($E_x$), Im($E_y$), and Re($H_z$) develops a well-defined grating-type electric-dipole pattern. This redistribution of multipolar content shows that the role of the grating extends beyond simply shifting the resonance frequency: it reorganizes the hybridization between the constituent multipoles. The well-defined $\mathrm{ED_y}$ field pattern indicates that this contribution has evolved from a weak background response into a coherent resonant displacement-current distribution. This reorganization is accompanied by reduced radiative coupling of the $\mathrm{hMD_x}$ hybrid mode, leading to a narrower spectral response and a higher quality factor.

The $\mathrm{hED_y}$ mode in Figure S6(b) exhibits contrasting behavior. Although signatures of both constituent resonances remain visible, the $\mathrm{MD_x}$ contribution is reduced toward a weak background response compared with the grating-free case in Figure S4(a). These results demonstrate that the grating does more than shift the underlying resonance frequencies; it actively reorganizes the multipolar composition of the collective eigenmodes. By selectively perturbing the constituent resonances according to their spatial overlap with the interconnection region of the grating, the geometry provides control over both their spectral separation and hybridization strength.

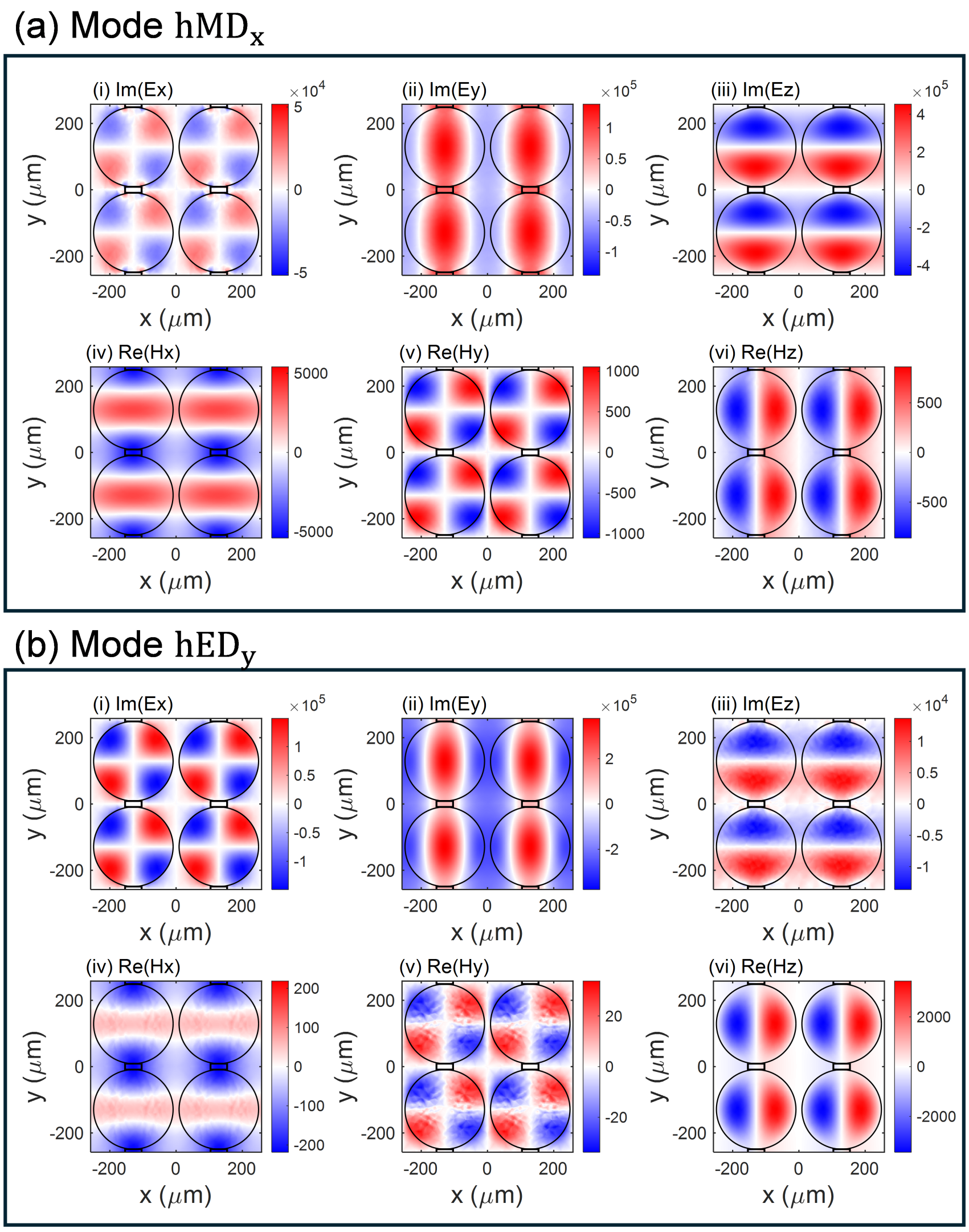

Figure S6: **Effects of grating on the multipolar composition of the hybrid modes** $\mathrm{hMD_x}$ **and** $\mathrm{hED_y}$ **under TE-polarized normal incidence (**$\mathbf{k}_z, \mathbf{E}_y, \mathbf{H}_x$**).** (a) Mode profile of the $\mathrm{hMD_x}$ mode. Introducing the grating strengthens the magnetic-dipole contribution $\mathrm{MD_x}$, as evidenced by Re($H_x$), Re($H_y$), and Im($E_z$). The field components Im($E_x$), Im($E_y$), and Re($H_z$) reveal a weaker but well-defined $\mathrm{ED_y}$ contribution, reflecting the grating-induced reorganization of the multipolar hybridization. (b) Mode profile of the $\mathrm{hED_y}$ mode, revealing its hybrid nature despite the $\mathrm{ED_y}$ contribution being more than one order of magnitude stronger than the $\mathrm{MD_x}$ contribution.

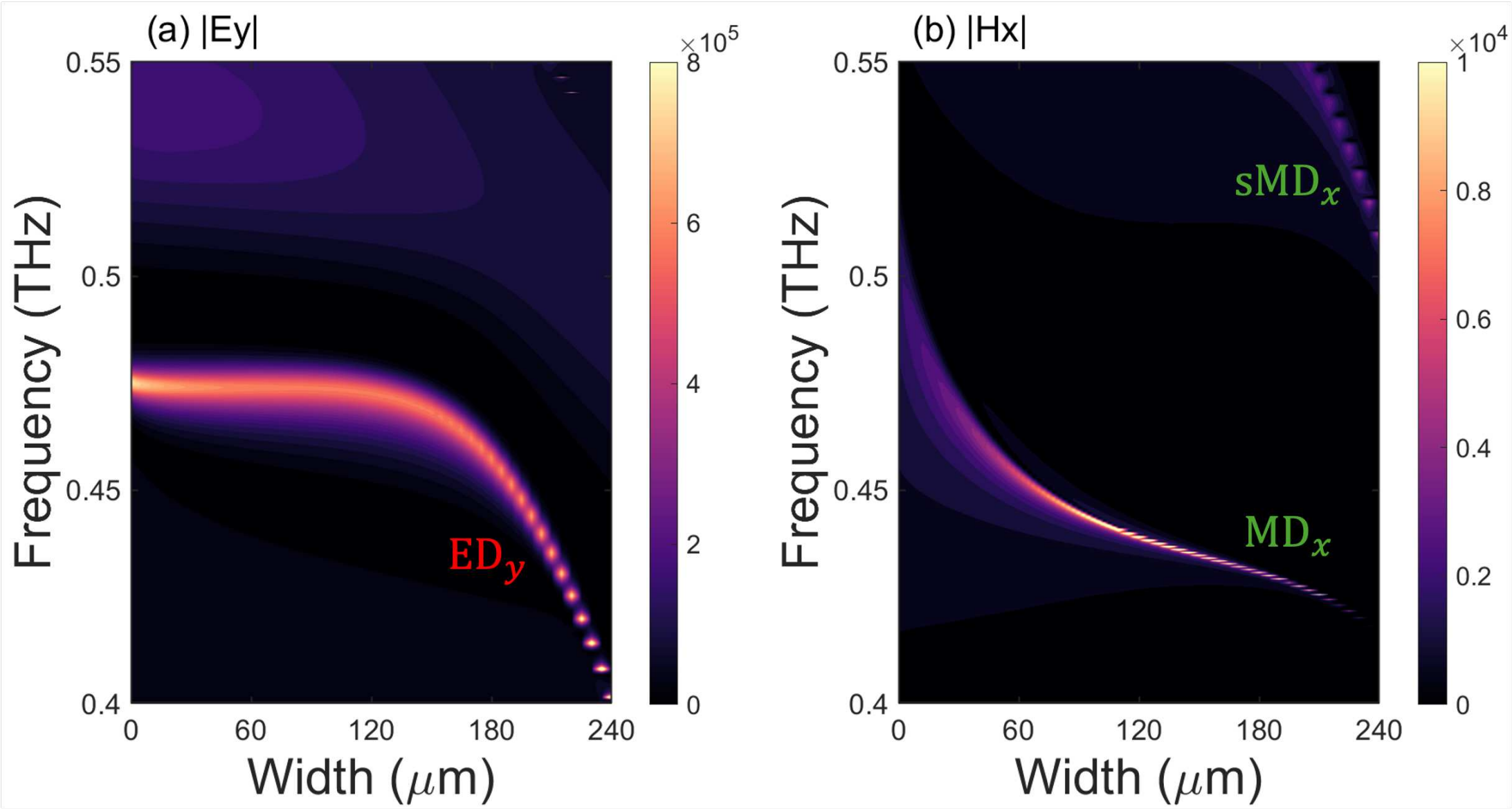

Figure S7: Spectral evolution of the three labeled modes, $\mathrm{ED_y}$, $\mathrm{MD_x}$, and $\mathrm{sMD}_x$, under TE-polarized normal incidence $(\mathbf{k}_z, \mathbf{E}_y, \mathbf{H}_x)$ as a function of grating width. The maximum width, $W = 240\ \mu$m, corresponds to the perfect-grating limit. (a) The $\mathrm{ED_y}$ resonance frequency remains nearly unchanged over the range $W = 0$–$130\ \mu$m because its displacement-current circulation lies predominantly in the $xz$ plane, resulting in weak spatial overlap with the $y$-oriented grating. For larger grating widths, the spatial overlap progressively increases as the hybrid metasurface evolves toward a continuous grating, leading to a stronger spectral shift of the $\mathrm{ED_y}$ resonance. At $W = 240\ \mu$m, the structure reaches the perfect-grating limit. (b) In contrast, the $\mathrm{MD_x}$ resonance undergoes a pronounced red shift already within the range $W = 0$–$130\ \mu$m because its displacement-current circulation lies predominantly in the $yz$ plane, resulting in strong spatial overlap with the interconnect region between the cylinder resonators. This strong sensitivity to the grating width results in a spectral crossing with the $\mathrm{ED_y}$ resonance. As w approaches $240\ \mu$m, the $\mathrm{MD_x}$ resonance disappears as the structure approaches continuous translational invariance along the $y$ direction. By contrast, the $\mathrm{sMD_x}$ resonance persists toward the perfect-grating limit because its field is predominantly confined in the transverse $xz$ plane and therefore remains compatible with the discrete periodicity retained along the $x$ direction.

Figure S7 shows the spectral evolution of the $\mathrm{ED_y}$ and $\mathrm{MD_x}$ resonances with grating width. As discussed above, the $\mathrm{ED_y}$ resonance remains nearly unchanged over the smaller grating-width range, $W = 0$–$130$ $\mu$m, because of its weak spatial overlap with the $y$-oriented grating, whereas the $\mathrm{MD_x}$ resonance undergoes a pronounced red shift owing to its strong overlap with the interconnect region. The distinct spectral responses of the two resonances therefore lead to their grating-induced spectral crossing.

The concurrent narrowing of the $\mathrm{MD_x}$ resonance with increasing grating width arises from reduced radiative coupling. In the grating-free or narrow-grating metasurface, the $\mathrm{MD_x}$ resonance couples through the lattice resonance to a strongly radiative electric-dipole contribution, resulting in a relatively broad spectral response [Figures S2 and S4(b)]. As the grating width increases, the $\mathrm{MD_x}$ resonance red-shifts while its coupling to the electric-dipole channel is reorganized into the more coherent resonant pattern shown in Figure S6(a). Simultaneously, the mode acquires an increasingly waveguide-like character, with stronger in-plane propagation and weaker coupling to the zeroth-order diffraction channel. Both effects reduce radiative loss, thereby increasing the quality factor and narrowing the resonance.

Over the wider grating-width range, $W = 130$–$240$ $\mu$m, the hybrid structure progressively evolves toward the perfect-grating limit. In this regime, the spatial overlap between the interconnection and the $\mathrm{ED_y}$ mode progressively increases, producing the pronounced red shift observed in Figure S7(a). Because the $\mathrm{ED_y}$ mode is predominantly confined within the transverse $xz$ plane, it persists as w approaches $240$ $\mu$m. In contrast, the $\mathrm{MD_x}$ resonance disappears toward the grating limit because its existence relies on field confinement, and correspondingly on structural discreteness, along the $y$ direction [Figure S4(b,iv)]. This confinement is lost when the structure becomes continuously translationally invariant along $y$. The $\mathrm{sMD_x}$ resonance, however, persists because its confinement is primarily associated with the $x$ direction [Figure S3(d)], along which the grating remains periodically structured. Consequently, the $\mathrm{sMD_x}$ mode remains visible even at $W = 240$ $\mu$m in Figure S7(b). We discuss this grating limit further in Sec. S5, where we show how Mie resonances provide additional degrees of freedom for engineering nonlocal metasurfaces.

Next, we examine the mode profile of the $\mathrm{EQ_z}$ mode that forms the Dirac-like cone with $\mathrm{hMD_x}$ under TM-polarized oblique incidence. Figure S8 shows the six Cartesian field components. Im($E_z$), Re($H_x$), and Re($H_y$) exhibit the characteristic field symmetry of the dominant $\mathrm{EQ_z}$ mode, whereas Im($E_x$), Im($E_y$), and Re($H_z$) reveal a weaker in-plane electric-dipole contribution $\mathrm{ED_x}$. The $\mathrm{ED_x}$ contribution is symmetry-forbidden under direct excitation by the incident plane wave and is instead induced through lattice-mediated coupling.

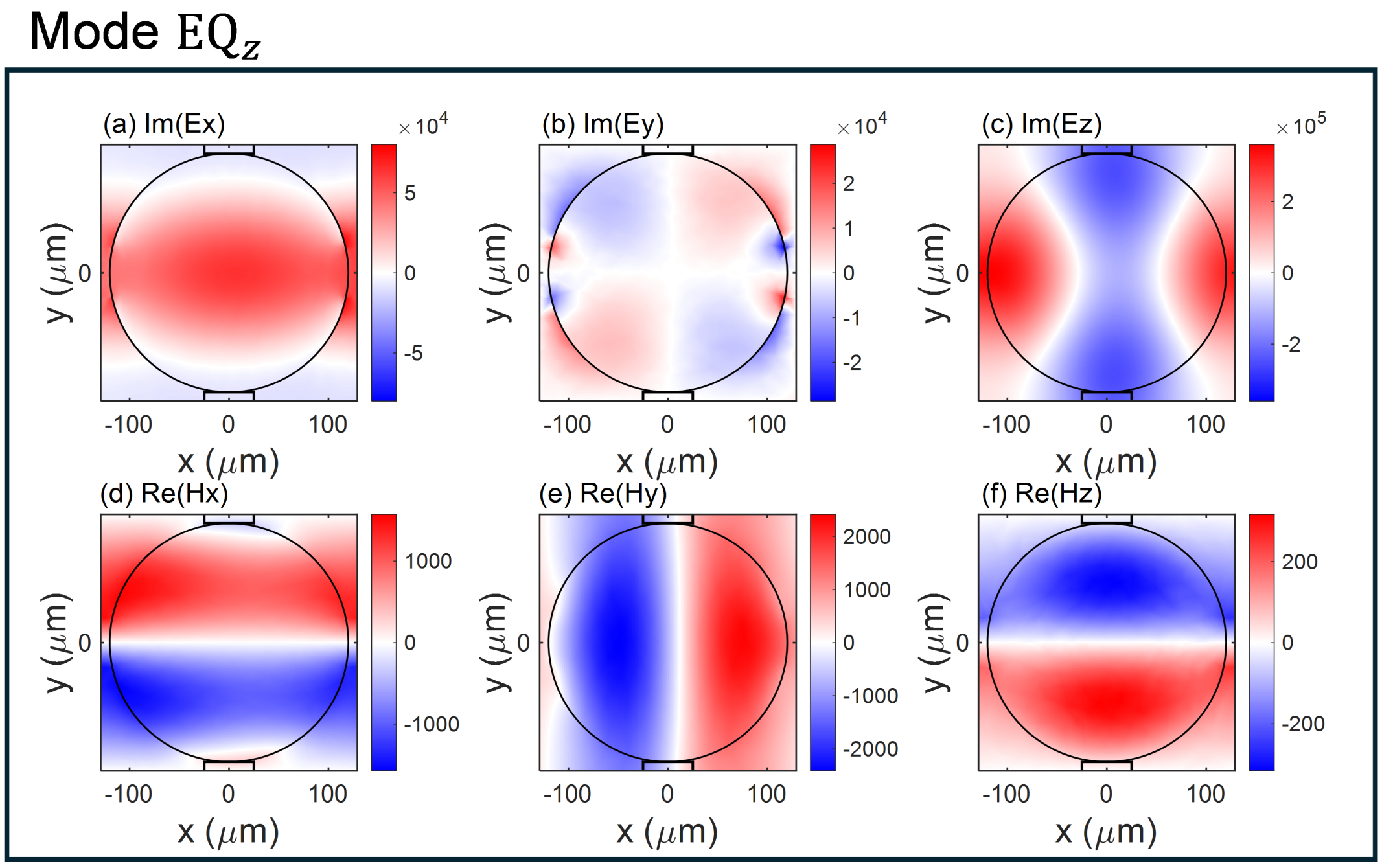

Figure S8: **Multipolar analysis of the electric-quadrupole mode** $\mathrm{EQ_z}$ **under TM-polarized oblique incidence at** $\theta_y = 2°$**.** The field patterns of Im($E_x$), Im($E_y$), and Re($H_z$) reveal an $x$-oriented electric-dipole contribution $\mathrm{ED_x}$. Im($E_z$), Re($H_x$), and Re($H_y$) exhibit the characteristic field symmetry of the dominant electric-quadrupole mode $\mathrm{EQ_z}$. The $\mathrm{ED_x}$ contribution is symmetry-forbidden from coupling directly to the incident plane wave and is instead induced through coupling to the lattice resonance.

In contrast to the dipolar $\mathrm{MD_x}$ resonance, the higher-order $\mathrm{EQ_z}$ mode exhibits a more spatially structured field distribution, with multiple field lobes distributed around the resonator boundaries. Moreover, the in-plane field symmetry of the $\mathrm{EQ_z}$ mode exhibits nodes near the inter-unit-cell gaps, as evidenced by the nearly vanishing Re($H_x$) and Re($H_y$) fields in the interconnection region [Figures S8(d,e)]. Consequently, the $\mathrm{EQ_z}$ mode has weaker spatial overlap with the interconnection and is therefore less sensitive to variations in grating width than the $\mathrm{MD_x}$ resonance. This contrast in grating sensitivity allows the strongly red-shifting $\mathrm{MD_x}$ band to approach the comparatively weakly perturbed $\mathrm{EQ_z}$ band, progressively close their spectral gap, and form the Dirac-like cone discussed in the main text.

This grating-width-controlled modulation of resonance frequency, linewidth, multipolar composition, and relative spectral ordering provides a versatile route for engineering the hybrid modes $\mathrm{hMD_x}$ and $\mathrm{hED_y}$. The strong sensitivity of $\mathrm{hMD_x}$ to the grating width enables substantial control of its dispersion and linewidth, whereas the comparatively weak sensitivity of the higher-order $\mathrm{EQ_z}$ mode provides a relatively stable band against which $\mathrm{hMD_x}$ can be tuned to close the spectral gap. Together, these mechanisms provide the microscopic basis for realizing the quasi-flat bands and Dirac-like cone discussed in the main text.

## S4 Fabricated metasurfaces and grating-width-dependent band dispersions

Figure S9 shows optical images of the fabricated nonlocal metasurfaces with grating widths ranging from $W = 25$ $\mu$m to 100 $\mu$m. As w increases, the geometry evolves progressively from a weakly connected Mie-resonator array toward a more continuous grating-like structure. This controlled structural evolution experimentally implements the Mie–grating hybridization strategy discussed above, enabling systematic tuning of the multipolar hybridization and resulting

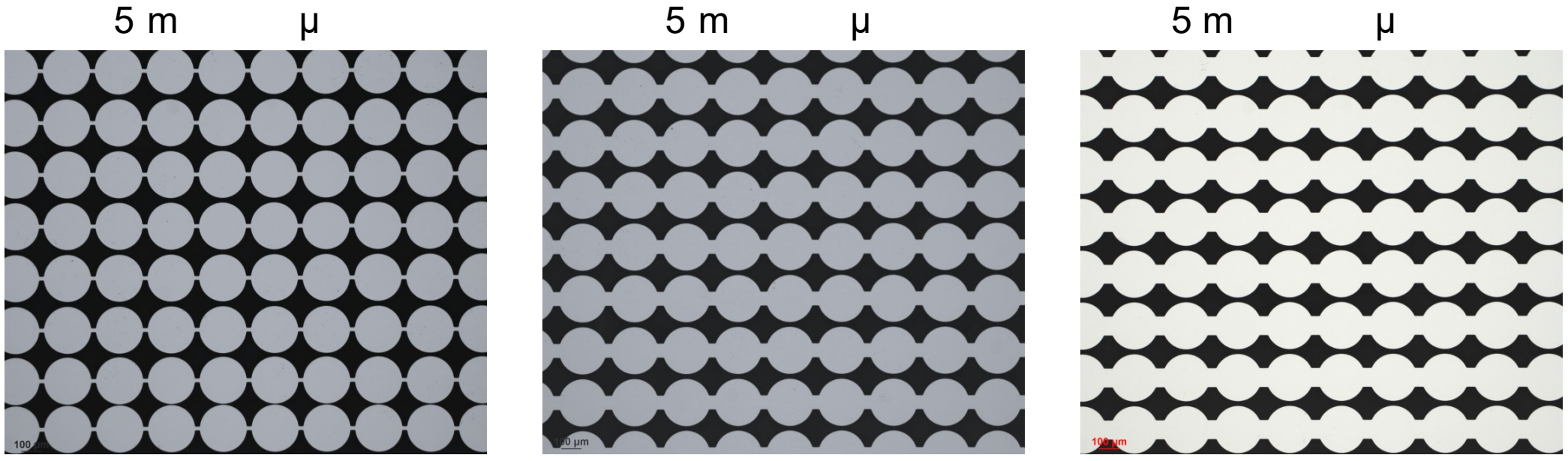

Figure S9: **Optical images of fabricated nonlocal metasurfaces with different grating widths w.** (a) $W = 25$ $\mu$m, (b) $W = 75$ $\mu$m, and (c) $W = 100$ $\mu$m. Increasing the grating width progressively evolves the metasurface geometry from an array of Mie resonators toward a grating-like structure, enabling controlled tuning of the multipolar hybridization and band dispersion.

band dispersions.

Figure S10 shows the corresponding simulated angle-resolved dispersions. Under TE-polarized incidence [Figures S10(a–c)], increasing the grating width strongly red-shifts the $\mathrm{hMD_x}$ band while simultaneously reducing its dispersion and spectral linewidth, leading to the formation of a quasi-flat band. Under TM-polarized incidence [Figures S10(d–f)], the same grating-induced red shift drives the $\mathrm{hMD_x}$ band toward the comparatively weakly perturbed $\mathrm{EQ_z}$ band, progressively closing their spectral gap and producing a Dirac-like cone. These width-dependent dispersions are consistent with the grating-controlled multipolar mechanism established above and agree well with the measured angle-resolved spectra presented in the main text.

# S5 Evolution from the hybrid Mie–grating regime to the perfect-grating limit

Figure S11 illustrates how the band structure evolves as the hybrid Mie–grating metasurface approaches the perfect-grating limit. Although a pure grating can support a rich variety of diffractive bands, including BICs, flat bands, and strongly dispersive photonic bands, restoring continuous translational invariance along the $y$ direction removes resonant degrees of freedom that rely on structural discreteness in that direction. This is evidenced by the disappearance of the $\mathrm{hMD_x}$, flat-band BIC (fBIC), and $\mathrm{EQ_z}$ bands at the grating limit. In the hybrid Mie–grating regime, the discrete resonators introduce additional Mie-type resonances that coexist and interact with grating-mediated resonances. Their hybridization therefore provides additional degrees of freedom for controlling spectral positions, linewidths, and interband coupling, enabling dispersions such as the quasi-flat $\mathrm{hMD_x}$ band and the Dirac-like cone that do not survive in the corresponding perfect-grating limit.

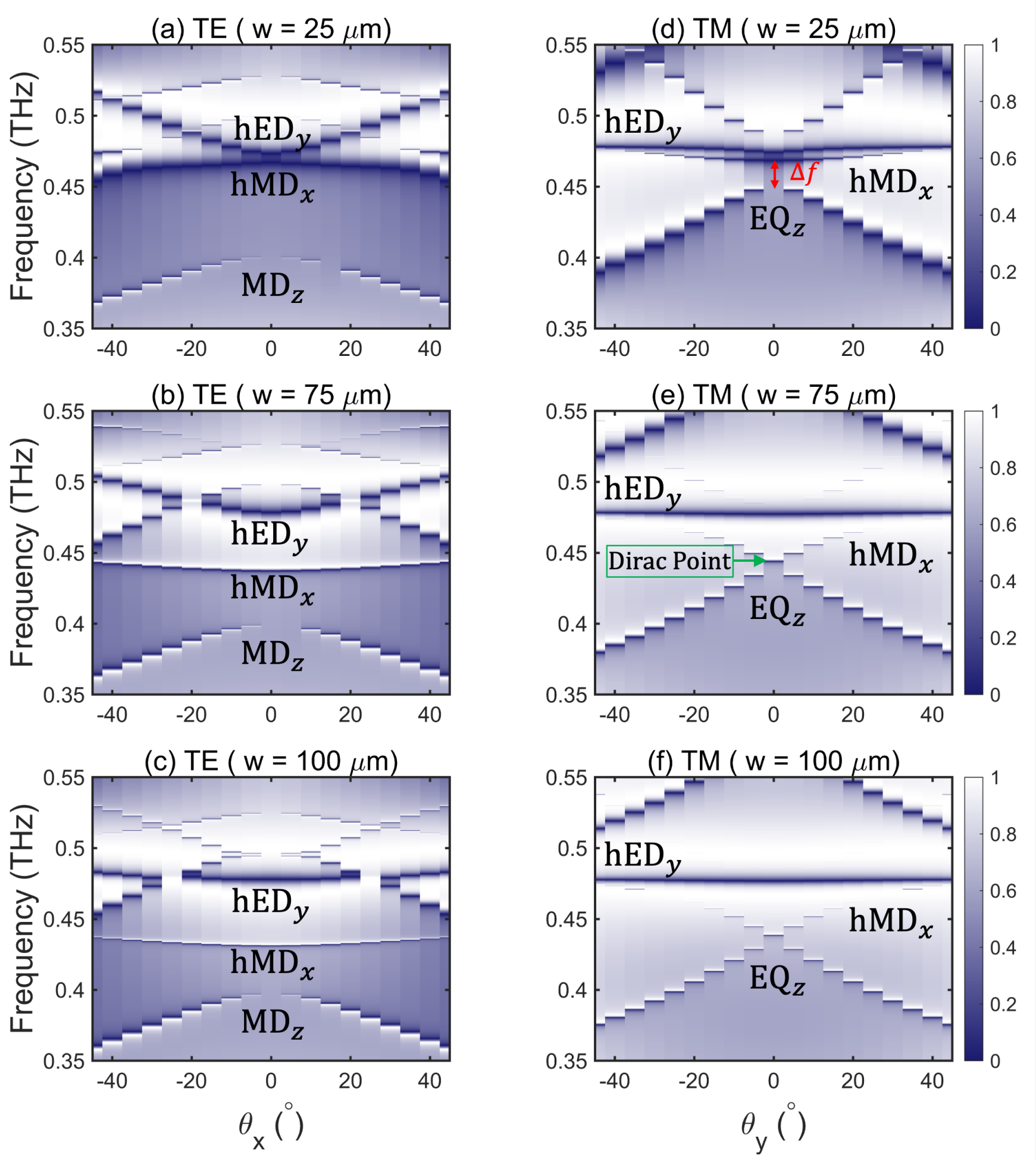

Figure S10: **Band-dispersion engineering with different grating widths w.** (a–c) Diffraction bands under TE-polarized incidence for grating widths $W = 25$ $\mu$m, 75 $\mu$m, and 100 $\mu$m, respectively. (d–f) Corresponding diffraction bands under TM-polarized incidence. Increasing the grating width progressively red-shifts the $\mathrm{hMD_x}$ band. This red shift closes the spectral gap $\Delta f$ between the $\mathrm{hMD_x}$ and $\mathrm{EQ_z}$ bands, leading to the formation of a Dirac-like cone. Under TE-polarized incidence, increasing the grating width simultaneously red-shifts, flattens, and narrows the $\mathrm{hMD_x}$ band, providing an effective route to realize a quasi-flat dispersion.

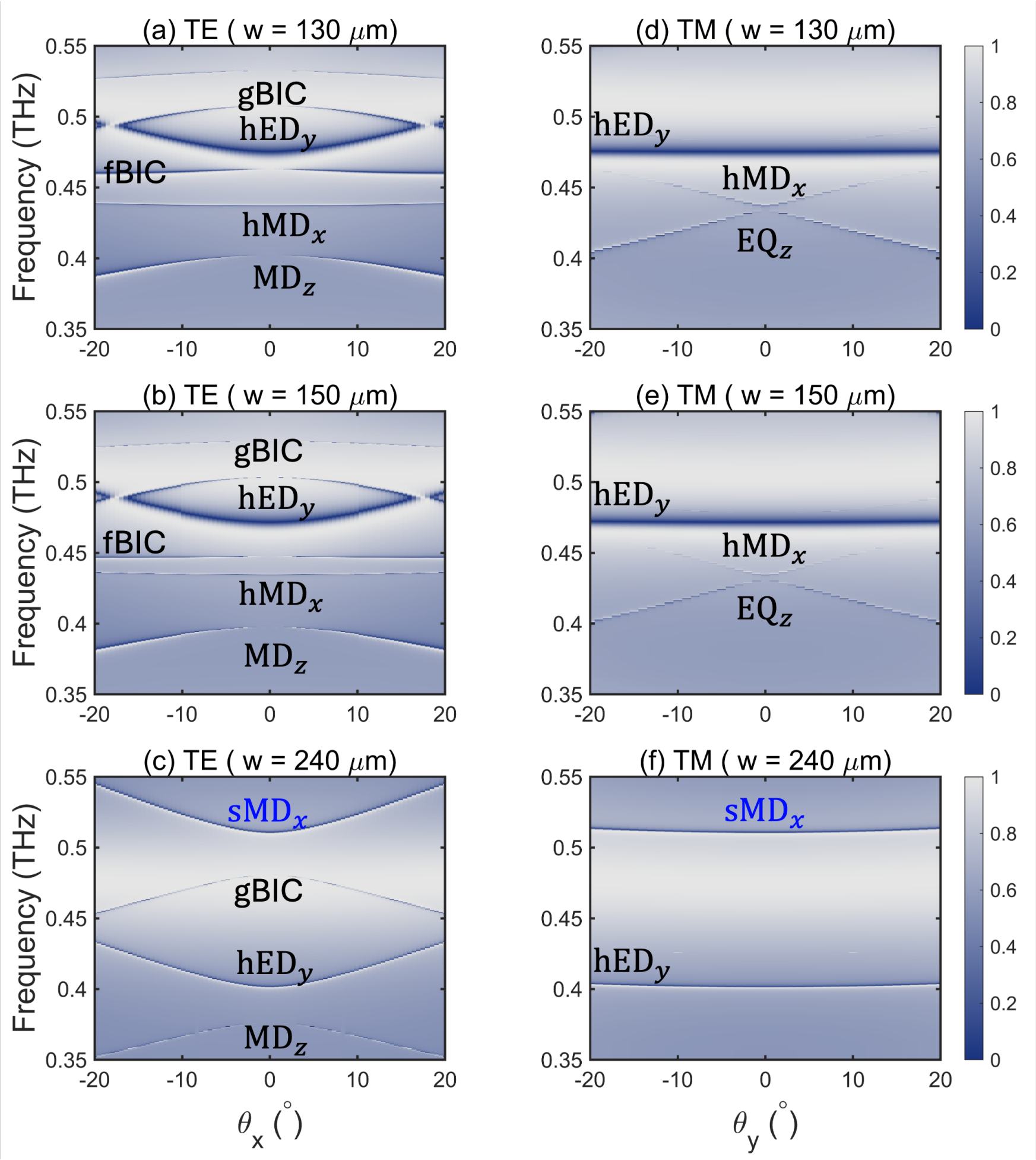

Figure S11: **Band-structure evolution from the hybrid Mie–grating regime to the perfect-grating limit.** (a–c) TE-polarized dispersions for grating widths $W = 130$ $\mu$m, 150 $\mu$m, and 240 $\mu$m, respectively. Increasing the grating width progressively transforms the hybrid Mie–grating metasurface into a continuous grating. As w approaches the perfect-grating limit at 240 $\mu$m, some bands, including the $\mathrm{hMD_x}$ band and the flat-band bound state in the continuum (fBIC), disappear because their existence relies on structural discreteness along the $y$ direction. Meanwhile, the $\mathrm{sMD_x}$ mode red-shifts into the frequency range of interest, as discussed in Figure S7, and gives rise to another quasi-flat band associated with grating diffraction. (d–f) Corresponding TM-polarized dispersions. At the perfect-grating limit, both the $\mathrm{hMD_x}$ and $\mathrm{EQ_z}$ bands disappear, together with the Dirac-like cone formed by their interaction. These results highlight the distinct roles of Mie and grating resonances in shaping the band structure of the hybrid metasurface.

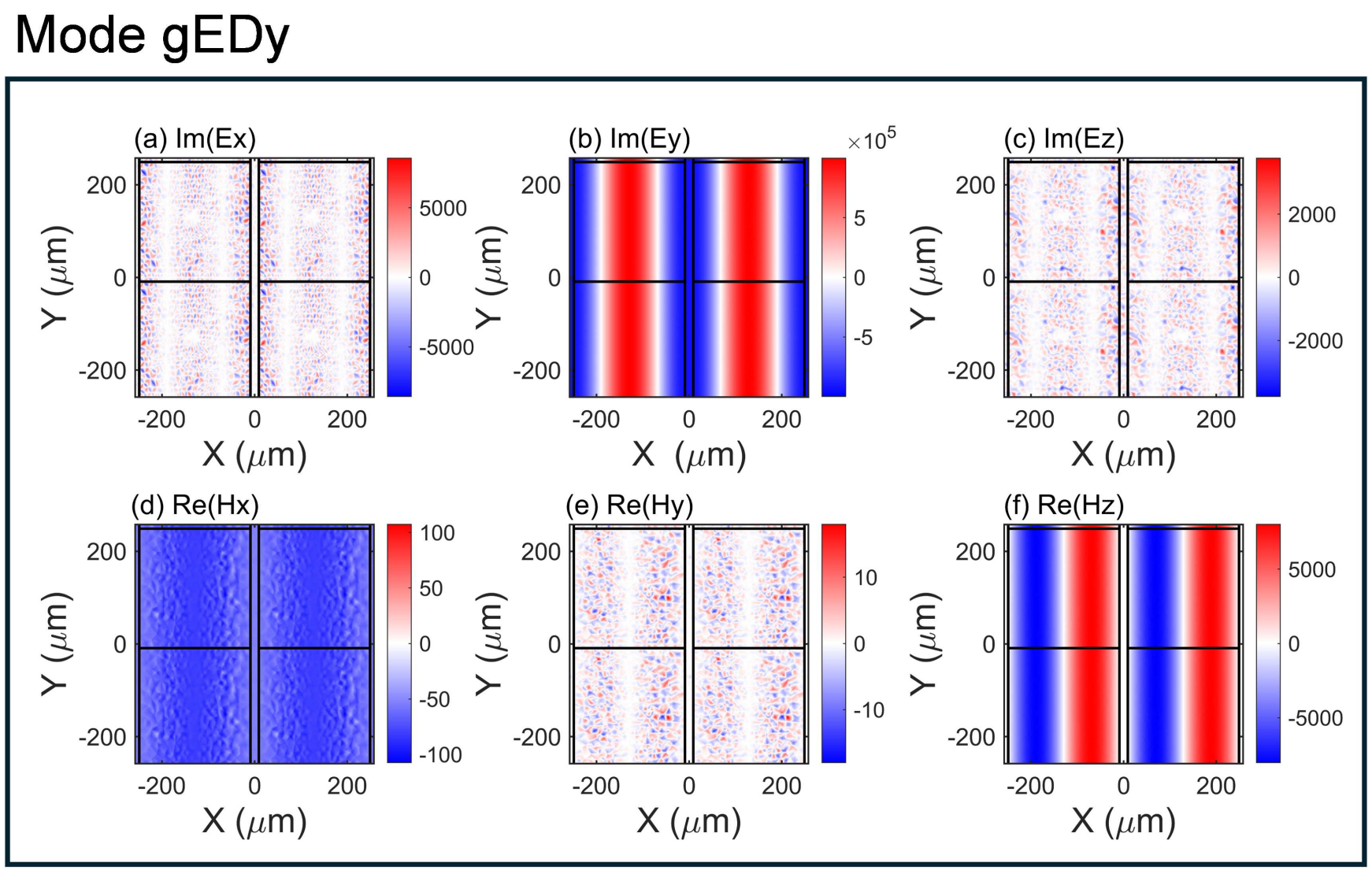

Figure S12: **Mode profile of the $\mathrm{hED_y}$ mode at the perfect-grating limit.** The field components Re($H_x$), Re($H_y$), and Im($E_z$), which characterize the $\mathrm{MD_x}$ contribution in the hybrid metasurface, become negligible at the grating limit, indicating the disappearance of the $\mathrm{MD_x}$ resonance. In contrast, Im($E_x$), Im($E_y$), and Re($H_z$) reveal the evolution of the $\mathrm{hED_y}$ mode in Figure S6 into a grating-type electric-dipole mode $\mathrm{gED_y}$. At the perfect-grating limit, the response is dominated by Im($E_y$) and Re($H_z$), consistent with the $y$-polarized excitation and the continuous translational invariance along the $y$ direction.

Figure S12 provides a microscopic view of this transition by examining the $\mathrm{hED_y}$ mode at the perfect-grating limit. As continuous translational invariance is restored along the $y$ direction, the $\mathrm{MD_x}$ contribution associated with the discrete resonator geometry vanishes, and the hybrid $\mathrm{hED_y}$ mode evolves into a predominantly grating-type electric-dipole mode $\mathrm{gED_y}$. The surviving response is dominated by the field components compatible with the grating symmetry and the incident polarization. This evolution highlights the complementary roles of Mie and grating resonances: grating diffraction alone can generate rich collective band structures, whereas breaking the continuous translational invariance through discrete resonators introduces additional localized resonant degrees of freedom. Hybridizing these Mie and grating resonances substantially expands the accessible modal landscape, providing additional degrees of freedom for engineering nonlocal metasurfaces beyond those offered by the purely diffractive grating limit.